\documentclass[letterpaper,journal]{IEEEtran}

\makeatletter
\newcommand{\iftwocolumn}[2]{%
  \if@twocolumn#1\else#2\fi
}
\makeatother

\usepackage{amsmath,amssymb}
\usepackage{dsfont}
\usepackage{algpseudocode}
\usepackage{algorithm}
\usepackage{array}
\usepackage[caption=false,font=normalsize,labelfont=sf,textfont=sf]{subfig}
\usepackage{textcomp}
\usepackage{stfloats}
\usepackage{graphicx}
\usepackage[noadjust]{cite}
\usepackage{mathtools}
\usepackage{booktabs}
\usepackage[hidelinks]{hyperref}

\newcommand{\tablelinespread}[1][1.15]{\linespread{#1}\selectfont}

\newcommand{\term}[1]{\emph{#1}}
\newcommand{\algIndentParagraph}[2][\algorithmicindent]{\parbox[t]{\linewidth-#1}{#2\strut}}
\newcommand{\algIndent}{\hspace{\algorithmicindent}}

\newcommand{\supplementary}[1]{#1 (see the supplementary material or \cite{wu2026clipped})}
\newcommand{\suppPseudoCode}{\supplementary{App.~C}}
\newcommand{\suppFiguresA}{\supplementary{App.~D}}
\newcommand{\suppStability}{\supplementary{App.~E}}
\newcommand{\suppFiguresB}{\supplementary{App.~F}}

\newtheorem{theorem}{Theorem}
\newtheorem{problem}{Problem}
\newtheorem{remark}{Remark}
\newenvironment{proofof}[1]{\begin{IEEEproof}[Proof of #1]}{\end{IEEEproof}}

\newcommand{\proofref}[1]{\leavevmode\unskip\hfill\penalty0\hbox{\quad}\nolinebreak\hfill\hbox{(Proof in #1.)}}

\newcommand{\supported}{\ensuremath{\checkmark}}
\newcommand{\notsupported}{\ensuremath{\times}}
\newcommand{\parameters}{\#Params}
\newcommand{\optimality}{\ensuremath{\star}Opt}
\newcommand{\nolessthan}[1]{\ensuremath{#1\textsuperscript{+}}}
\newcommand{\nomorethan}[1]{\ensuremath{{\le}#1}}

\newcommand{\eqdef}{:=}
\newcommand{\geqdef}{\triangleq}
\newcommand{\relvar}[2]{\buildrel \text{#1} \over #2}
\newcommand{\eqvar}[1]{\relvar{#1}{=}}

\newcommand{\eqdist}{\eqvar{\text{d}}}

\newcommand{\real}{\mathbb{R}}
\newcommand{\en}{\mathrm{e}}

\DeclareMathOperator{\sigmoid}{sigmoid}
\DeclareMathOperator{\softplus}{softplus}
\DeclareMathOperator{\bin}{bin}
\DeclareMathOperator{\mlp}{mlp}

\DeclarePairedDelimiterX{\clip}[3]{\langle}{\rangle_{#2,#3}}{#1}
\DeclarePairedDelimiterX{\lowerclip}[2]{\langle}{\rangle_{\ge #2}}{#1}
\DeclarePairedDelimiterX{\upperclip}[2]{\langle}{\rangle_{\le #2}}{#1}

\DeclarePairedDelimiterX{\floor}[1]{\lfloor}{\rfloor}{#1}

\DeclarePairedDelimiter{\set}{\{}{\}}

\newcommand{\indicator}{\mathds{1}}
\newcommand{\diff}{\mathrm{d}}
\newcommand{\expect}{\mathbb{E}}
\newcommand{\borel}{\mathfrak{B}}

\newcommand{\measure}[1]{\mathfrak{#1}}
\newcommand{\onepoint}[1]{\delta_{#1}}
\newcommand{\bernoulli}{\measure{B}}
\newcommand{\expon}{\measure{E}}
\newcommand{\uniform}{\measure{U}}

\newcommand{\clippedmean}{\bar{\mu}}
\newcommand{\mcratio}{\bar{\rho}}

\DeclareMathOperator{\averageReward}{\mathcal{G}}

\newcommand{\chargeEfficiency}{\eta_{\text{c}}}
\newcommand{\dischargeEfficiency}{\eta_{\text{d}}}
\newcommand{\maxChargeEnergy}{E_{\text{cmax}}}
\newcommand{\maxDischargeEnergy}{E_{\text{dmax}}}

\newcommand{\reserve}[1]{\overline{#1}}
\newcommand{\parameterizedPolicy}[3][]{%
  #2_{\text{#3}%
  \ifx%
    \\#1\\%
  \else%
    (#1)%
  \fi}%
}
\newcommand{\clippedGreedyPolicy}[1][]{\parameterizedPolicy[#1]{\sigma}{cg}}
\newcommand{\maximinOptimalPolicy}[1][]{\parameterizedPolicy[#1]{\sigma}{mo}}
\newcommand{\fixedFractionPolicy}[1][]{\parameterizedPolicy[#1]{\sigma}{ff}}
\newcommand{\optimisticClippedAffinePolicy}[1][]{\parameterizedPolicy[#1]{\sigma}{oca}}
\newcommand{\robustClippedAffinePolicy}[1][]{\parameterizedPolicy[#1]{\sigma}{rca}}

\newcommand{\ocaOl}[1][]{\parameterizedPolicy[#1]{\sigma}{oca-ol}}
\newcommand{\rcaOl}[1][]{\parameterizedPolicy[#1]{\sigma}{rca-ol}}

\newcommand{\maximinOptimalLinear}{s_{\text{mol}}}
\newcommand{\maximinOptimalLinearHat}{\hat{s}_{\text{mol}}}

\newcommand{\lookahead}[1]{\dot{#1}}

\begin{document}

\title{Clipped Affine Policy: Low-Complexity Near-Optimal Online Power Control for Energy Harvesting Communications over Fading Channels}

\author{%
  Hao~Wu, Shengtian~Yang, Huiguo~Gao, Diao~Wang, Jun~Chen, Guanding~Yu%
  \thanks{A conference paper containing part of this work was presented at the 2026 IEEE International Symposium on Information Theory~\cite{wu2026robust}.}
  \thanks{Corresponding Author: Shengtian Yang.}%
  \thanks{Hao~Wu and Shengtian~Yang are with the School of Information and Electronic Engineering (Sussex Artificial Intelligence Institute), Zhejiang Gongshang University, Hangzhou 310018, China (e-mail: \mbox{wu\_hao\_a@126.com}; \mbox{yangst@codlab.net}.)}%
  \thanks{Huiguo~Gao, Diao~Wang, and Guanding~Yu are with the College of Information Science and Electronic Engineering, Zhejiang University, Hangzhou 310027, China (e-mail: \mbox{huiguogao@zju.edu.cn}; \mbox{diorwang@zju.edu.cn}; \mbox{yuguanding@zju.edu.cn})}
  \thanks{Jun~Chen is with the Department of Electrical and Computer Engineering, McMaster University, Hamilton, ON L8S 4K1, Canada (e-mail: \mbox{chenjun@mcmaster.ca}).}%
}



\maketitle

\begin{abstract}
  This paper studies online power control for battery-limited point-to-point energy harvesting communications over slow block-fading channels.
  A linear-policy-based approximation is developed for the relative-value function in the Bellman equation of the power control problem.
  This approximation leads to two fundamental parameterized clipped affine policies: an optimistic policy derived from a certainty-equivalence-type approximation and a robust policy derived from worst-case analysis.
  For independent and identically distributed energy arrivals and channel states, two families of power control schemes are developed based on the optimistic clipped affine (OCA) and robust clipped affine (RCA) policies, respectively.
  The proposed adaptive RCA policy based on reinforcement learning (RCA-RL) is further extended to address four scenarios with contextual information: one-step energy lookahead, one-step channel lookahead, one-step joint energy-channel lookahead, and Markov energy arrivals.
  Extensive simulation results show that the proposed schemes provide a favorable tradeoff between computational complexity and performance.
  The adaptive RCA policy based on the maximin optimal linear-policy-slope approximation (RCA-OLA-A) and the RCA-RL scheme achieve the best overall performance, while the RCA policy based on the maximin optimal linear policy (RCA-OL) is the best-performing closed-form policy.
  In particular, RCA-OLA-A, RCA-RL, and the aforementioned RCA-RL extensions achieve less than 2\% performance loss relative to the optimal policy across a range of scenarios, consistently outperforming the considered benchmark approaches, including generic reinforcement learning baselines.
  The RCA-OL policy also performs well with less than 4\% performance loss.
\end{abstract}


\section{Introduction}

Recent advances in energy harvesting (EH) technologies enable self-sustaining wireless communication by allowing devices to replenish energy from ambient sources (e.g., solar, wind, or radio-frequency), reducing maintenance and extending operational lifetime.
This capability is particularly attractive for Internet-of-Things (IoT) deployments such as environmental monitoring, surveillance, and safety-critical sensing, where large numbers of low-power nodes must operate for long periods with limited human intervention.
However, the harvested energy supply is inherently intermittent and stochastic, which makes transmit power control substantially more challenging than in conventional communication systems with stable energy supplies.
There has been a large body of literature on this topic, e.g.,~\cite{ulukus2015energy,ku2016advances,ma2020sensing,hu2020modeling,alamu2025machine,yang2025power} and the references therein.
In this paper, we study online power control for battery-limited point-to-point EH communications over wireless fading channels (or simply, point-to-point EH communication power control), with or without lookahead information.
The utility of interest is typically throughput-based, but it may also be any increasing concave utility function of the energy allocation.

A special case of this problem is the quasi-static fading scenario, where the channel signal-to-noise ratio (SNR) coefficient remains constant throughout the entire transmission duration.
The optimal policy for this case was derived under Bernoulli energy arrivals with or without energy lookahead~\cite{kazerouni2015optimal,shaviv2015capacity,shaviv2016universally,yang2020maximin,zibaeenejad2022optimal}.
For general independent and identically distributed (i.i.d.) energy arrivals, previous works focus on the greedy policy, which is optimal in the low-battery-capacity regime~\cite{wang2021optimality}, and universally near-optimal policies, including the fixed fraction policy~\cite{shaviv2016universally}, the maximin optimal policy~\cite{yang2020.6maximin,yang2020maximin}, the locally fixed fraction policy~\cite{yang2020.6maximin,yang2025power}, the maximin optimal linear policy~\cite{garmaroudi2022linear}, and the two-piece fixed fraction policy~\cite{yang2025power}.
In the case of non-i.i.d.\ energy arrivals, no satisfactory closed-form solution is available.

Although the quasi-static fading scenario is relatively simple, it already reveals two structural properties and one design principle underlying good policies:
\begin{itemize}
  \item \textbf{Worst-case optimality implies near-optimality}:
  For a family of i.i.d.\ energy-arrival distributions with the same clipped mean, any policy that is optimal or near-optimal in the worst case over this family, and satisfies certain mild conditions, is near-optimal for every distribution in the family~\cite{shaviv2016universally,arafa2018online,yang2020maximin,yang2025power}.

  \item \textbf{Linear-policy structure}:
  With an appropriately chosen slope, a linear policy can achieve optimal or near-optimal performance~\cite{shaviv2016universally,arafa2018online,wang2021optimality,garmaroudi2022linear,yang2025power}.

  \item \textbf{Worst-case analysis as a design principle}:
  Worst-case analysis provides a principled way to construct universally near-optimal policies~\cite{shaviv2016universally,arafa2018online,yang2020maximin,garmaroudi2022linear,yang2025power}.
\end{itemize}

In the case of slow block-fading scenario, where the channel SNR coefficient remains constant within each time slot but varies across different time slots, the problem is more challenging due to the two-dimensional variation of the energy arrivals and the channel SNR coefficients.
Except for some special cases, e.g., independent Bernoulli energy arrivals and channel SNR coefficients~\cite{khajepour2021optimal}, the optimal policy is not known in general.
Most research focuses on reinforcement learning (RL)-based power control designs (e.g., \cite{ortiz2016reinforcement,masadeh2018reinforcement,kim2018actionbounding,li2019deep,masadeh2019actorcritic,qiu2019deep,kim2021shallow}), which aim to learn optimal policies from data.
However, the schemes proposed in these studies often entail substantial computational complexity, and the optimality of the learned policies lacks rigorous verification.
Moreover, most of these studies focus only on power control with one-step energy lookahead, which is easier to solve than online power control without lookahead information.
This is because ideas from optimal offline power control (e.g., \cite{ozel2011transmission}) remain effective to some extent in the lookahead setting.
The assumption of perfect one-step energy lookahead is also overly idealized.

Because of the theoretical difficulty of analyzing the point-to-point EH communication power control problem, relatively little work has been devoted to it in recent years.
Moreover, the insights and methods developed for the quasi-static fading version of this problem, as well as its extensions to certain multiuser settings (e.g., multiple-access and broadcast channels~\cite{baknina2016optimal,baknina2018energy}), have largely been overlooked in recent literature and have not been carried over to more complex problems with similar underlying structures, including the slow block-fading version considered here.
Most recent work instead focuses on more complex resource-allocation problems in energy-harvesting systems, such as multi-task resource allocation~\cite{yao2023multiple}, joint device scheduling and power control for federated learning~\cite{zhang2026federated}, and energy-efficient power management in small-cell networks~\cite{cho2025multiagent}.
However, the methods used to solve these more complicated problems are mostly generic numerical approaches for Markov decision processes (MDPs), such as value iteration, and generic RL methods.
These numerical approaches are computationally expensive and provide little insight into the structure of the optimal policy.
To improve the performance of RL-based designs, only simple properties, such as monotonicity~\cite{kim2021shallow,zhang2026federated} and non-optimal-action region~\cite{kim2018actionbounding}, have been exploited.

As a result, the point-to-point EH communication power control problem in the slow block-fading case is often regarded as an isolated, outdated problem of limited interest.
It is also often regarded as practically resolved by RL-based approaches and by certain suboptimal methods, such as Lyapunov optimization~\cite{amirnavaei2016online} and prediction methods based on the optimal offline policy~\cite{ku2021neuralnetworkbased}.
However, the problem remains fundamentally open in the sense that no general closed-form optimal solution or fundamental structure of the optimal policy is known, let alone its connection to more complex problems.
This impasse arises from the challenge of solving the Bellman equation for an EH communication power control problem formulated as an MDP.
Consequently, it is difficult to obtain a tractable structural characterization of the optimal policy and its associated relative-value function.
While RL combined with neural networks offers a universal, out-of-the-box approach to empirically solving the Bellman equation, it is important to note three limitations: (i) the optimality of learned policies produced by deep RL methods is rarely rigorously verified, (ii) RL-based approaches are often computationally expensive, and (iii) their stability and convergence are not guaranteed, particularly because of sensitivity to hyperparameters.
These limitations can be mitigated by a deeper understanding of the structure of the optimal policy and its associated relative-value function.

To address this impasse, we adopt an approximation-based approach.
By examining the approximate structure of the relative-value function associated with the Bellman equation underlying the power control problem, we successfully extend the insights and methods developed for the quasi-static fading case to the slow block-fading case.
This extension leads to a linear-policy-based approximation of the relative-value function and a collection of low-complexity, near-optimal policies.

The main contributions of this paper are as follows:
\begin{enumerate}
  \item We find a linear-policy-based approximation (Eq.~\eqref{eq:relative_value_function_approximation2}) to the relative-value function in the Bellman equation of the power control problem.
  Based on this approximation, we derive two parameterized policies using a certainty-equivalence-type approximation and worst-case analysis, respectively, corresponding to Eqs.~\eqref{eq:optimistic_clipped_affine_policy} and~\eqref{eq:robust_clipped_affine_policy}.
  Both take the form
  \begin{equation}
    \sigma(b,\gamma)
    = \clip*{\theta_0 + \theta_1 b - \theta_2 \frac{1}{\gamma}}{b_0}{b_1},\ \theta_0,\theta_1,\theta_2 \ge 0,
  \end{equation}
  where $b$ and $\gamma$ denote the battery level and channel SNR coefficient, respectively, $\clip{x}{z_0}{z_1} \geqdef \min\{\max\{x,z_0\},z_1\}$ clips $x$ to $[z_0,z_1]$, and $b_0$, $b_1$ are battery-level-dependent bounds satisfying $0 \le b_0 \le b_1 \le b$.
  These policies are thus coined \term{clipped affine policies}, distinguished as \term{optimistic} and \term{robust} ones.
  Their derivations (Eqs.~\eqref{eq:optimistic_waterfilling} and~\eqref{eq:robust_waterfilling}) reveal a battery-limited weighted directional waterfilling mechanism operating between adjacent time slots, an online counterpart of the directional waterfilling principle~\cite{ozel2011transmission} in the offline setting.

  \item Building on the relative-value approximation and the derived policies, we propose two families of power control schemes for i.i.d.\ energy arrivals and channel states: those based on the optimistic clipped affine (OCA) policy and those based on the robust clipped affine (RCA) policy.
  Based on the physical interpretation of key parameters, we further extend the RL-based adaptive RCA policy (RCA-RL) to address four scenarios with contextual information, including one-step energy lookahead, one-step channel lookahead, one-step joint energy-channel lookahead, and Markov energy arrivals.
  Extensive simulation results show that the proposed schemes provide a favorable tradeoff between computational complexity and performance.
  The adaptive RCA policy based on the maximin optimal linear-policy-slope approximation (RCA-OLA-A) and the RCA-RL scheme achieve the best overall performance, while the RCA policy based on the maximin optimal linear policy (RCA-OL) is the best-performing closed-form policy.
  In particular, under charging and discharging constraints, both the RCA-OLA-A and RCA-RL schemes incur less than about 1\% performance loss relative to the optimal policy across a range of scenarios, substantially outperforming the Lyapunov-optimization-based scheme~\cite{amirnavaei2016online}, a scheme especially designed for energy storage with charging and discharging constraints.
  Moreover, simulation results show that the RCA-RL scheme and its extensions for Markov energy arrivals significantly outperform generic RL baselines under both i.i.d.\ and wind-energy arrivals, the latter exhibiting strong temporal correlation.
  Table~\ref{tab:comparison} summarizes the proposed methods and existing approaches designed for the same or a similar EH communication model, highlighting that the proposed methods use only a few parameters while achieving very small performance loss relative to the optimal policy.
\end{enumerate}

\begin{table*}
  \centering
  \caption{Comparison of the Proposed Methods with Existing Approaches\textnormal{\textsuperscript{a}}}\label{tab:comparison}
  \tablelinespread[\iftwocolumn{1.05}{1.0}]
  \begin{tabular}{p{1.4cm}p{3.3cm}p{3cm}p{1.95cm}p{1.9cm}p{1.6cm}p{1.9cm}}
    \hline
    Paper & Method & Energy (E) and Channel (C) Models for Simulation & Online & Energy Lookahead                                                               & Channel Lookahead & Energy-Channel Lookahead \\
    \hline
    \cite{amirnavaei2016online}
    & Lyapunov optimization
    & E: Poisson-uniform compound\par C: Rayleigh
    & \supported\par   \parameters: 1\par \optimality: No\textsuperscript{b}
    & \notsupported
    & \notsupported
    & \notsupported \\
    \cite{ortiz2016reinforcement}
    & State-action-reward-state-action (SARSA) RL with $3$ binary features
    & E: Uniform\par C: Rayleigh
    & \notsupported
    & \supported\par \parameters: 3\par \optimality: \nomorethan{6\%}
    & \notsupported
    & \notsupported \\
    \cite{masadeh2018reinforcement}
    & SARSA RL
    & E: Binary Markov\par C: Binary Markov
    & \notsupported
    & \supported\par \parameters: \nolessthan{10} (for a $2$-unit-capacity battery) \par \optimality: \nomorethan{15\%}
    & \notsupported
    & \notsupported \\
    \cite[Alg.~2]{kim2018actionbounding}
    & Deep Q-Network (DQN) RL with action bounding
    & E: Gaussian\par C: Log-normal
    & \notsupported
    & \supported\par \parameters: \nolessthan{300}\par \optimality: \nomorethan{8\%}
    & \notsupported & \notsupported \\
    \cite{li2019deep}
    & Adaptive modulation based on DQN RL with a reward function derived from the modulation layer's bit rate, constrained by a target bit error rate.
    & E: Uniform\par C: Rayleigh
    & \supported\par \parameters: \nolessthan{100}\par \optimality: Unknown
    & \notsupported
    & \notsupported
    & \notsupported
    \\
    \cite{masadeh2019actorcritic}
    & Actor-critic RL
    & E: Gaussian random walk\par C: Gaussian random walk
    & \notsupported
    & \supported\par \parameters: \nolessthan{100}\par \optimality: Unknown
    & \notsupported
    & \notsupported \\
    \cite{qiu2019deep}
    & Deep Deterministic Policy Gradient (DDPG) RL with a net-bit-rate reward function
    & E: Solar\par C: Rayleigh
    & \notsupported
    & \supported\par \parameters: \nolessthan{17\text{k}}\par \optimality: Unknown
    & \notsupported
    & \notsupported \\
    \cite{kim2021shallow}
    & DPG RL with monotonic shape constraints, using generalized mutual information as the reward function
    & E: Bernoulli\par C: Rician
    & \notsupported
    & \supported\par \parameters: \nolessthan{20}\par \optimality: Unknown
    & \notsupported
    & \notsupported \\
    \cite{ku2021neuralnetworkbased}
    & [SU-GreenPCNet]: Prediction based on optimal offline policy and neural network
    & E: Solar\par C: Rayleigh
    & \supported\par \parameters: \nolessthan{30\text{k}}\par \optimality: \nomorethan{4.0\%}
    & \notsupported
    & \notsupported
    & \notsupported \\
    This paper
    & \textbf{RCA-OL}\par (Tabs.~\ref{tab:closed_form_policies}, \ref{tab:performance_loss}, and~\ref{tab:limited_performance_loss})
    & E: Bernoulli, exponential, uniform\par C: Rayleigh
    & \supported\par \parameters: 0\par \optimality: \nomorethan{3.7\%}
    & \notsupported
    & \notsupported
    & \notsupported \\
    This paper
    & \textbf{RCA-OLA-A}\par (Alg.~\ref{alg:rca-ola-a} and Tabs.~\ref{tab:performance_loss} and~\ref{tab:limited_performance_loss})
    & E: Bernoulli, exponential, uniform\par C: Rayleigh
    & \supported\par \parameters: 1\par \optimality: \nomorethan{\textbf{1.3\%}}
    & \notsupported
    & \notsupported
    & \notsupported \\
    This paper
    & \textbf{RCA-RL and its extensions}\par (Alg.~\ref{alg:rca-rl}, Sec.~\ref{sec:prediction_based_extensions}, and Tabs.~\ref{tab:performance_loss}, \ref{tab:limited_performance_loss}, \ref{tab:lookahead_performance_loss}, \ref{tab:generic_rl_comparison_under_uniform}, and \ref{tab:generic_rl_comparison_under_wind})
    & E: Bernoulli, exponential, uniform, wind\par C: Rayleigh
    & \supported\par \parameters: 4 (in the i.i.d.\ setting)\par \optimality: \nomorethan{1.5\%}\textsuperscript{c}
    & \supported\par \parameters: 3\par \optimality: \nomorethan{\textbf{1.8}\%}
    & \supported\par \parameters: 5\par \optimality: \nomorethan{\textbf{1.7\%}}
    & \supported\par \parameters: 4\par \optimality: Unknown \\
    \hline
  \end{tabular}
  \smallbreak
  \begin{minipage}{\textwidth}
    \footnotesize\rightskip=1em\hangindent=3.5em
    \makebox[3.5em]{\textit{Notes}:}%
    \textsuperscript{a}Key metrics for comparison: (i) parameter count (\parameters) and (ii) optimality (\optimality) in terms of performance loss relative to the optimal policy.\par
    \leftskip=3.5em
    \textsuperscript{b}Although~\cite{amirnavaei2016online} does not provide numerical optimality results, its derived policy exhibits a zero-output behavior below a battery threshold in the quasi-static-fading case, which differs markedly from known optimal policies (e.g.,~\cite{shaviv2016universally,wang2021optimality}).
    Our simulation results further indicate that it is far from optimal (see Tables~\ref{tab:performance_loss} and~\ref{tab:limited_performance_loss}).\par
    \textsuperscript{c}The optimality of two RCA-RL extensions (RCA-RL-M and RCA-RL-MP) under wind energy arrivals is unknown, because the optimal policy is not computable in this case.
    However, they significantly outperform the optimal policy under the first-order empirical distribution of wind energy arrivals, achieving more than 2\% performance gain.
  \end{minipage}
\end{table*}

The rest of this paper is organized as follows.
In Sec.~\ref{sec:problem_formulation}, we formulate the power control problem as an MDP.
In Sec.~\ref{sec:clipped_affine_policy}, we obtain an approximation for the relative-value function and use it to derive the clipped affine policies.
In Sec.~\ref{sec:practical_power_control}, we propose a collection of power control schemes based on clipped affine policies, where the RCA-RL scheme is further extended to scenarios with contextual information in Sec.~\ref{sec:prediction_based_extensions}.
Sec.~\ref{sec:simulation_results} presents the simulation results, and Sec.~\ref{sec:discussion_and_conclusion} concludes with a discussion.
For clarity and continuity, some proofs, pseudocode, and simulation results are given in the appendices, while Appendices C--F, which contain the omitted pseudocode and simulation results, are available either in the supplementary material associated with this paper or in the extended version of this paper~\cite{wu2026clipped}.
Table~\ref{tab:notation} summarizes the common notation used in this paper, and Table~\ref{tab:schemes} in Sec.~\ref{sec:simulation_results} lists the abbreviations of all schemes.

\begin{table}
  \centering
  \caption{Common Notation in This Paper}
  \label{tab:notation}
  \tablelinespread
  \begin{tabular}{lp{6.5cm}}
    \toprule
    Notation &
    Meaning \\
    \midrule
    $\geqdef$ &
    Global definition \\
    $\eqdef$ &
    Local definition \\
    $\eqdist$ &
    Equality in distribution \\
    $\onepoint{e}$ &
    One-point distribution at $e \in \real$ \\
    $\bernoulli_{q,e}$ &
    Bernoulli distribution $q\onepoint{e} + (1-q)\onepoint{0}$ \\
    $\bernoulli_q$ &
    $\bernoulli_{q,c}$ with $c$ denoting the battery capacity \\
    $\expon_\lambda$ &
    Exponential distribution with density $\lambda e^{-\lambda x}\indicator\set{x\ge 0}$ \\
    $\uniform_b$ &
    Uniform distribution over $[0,b]$ \\
    $\clip{x}{z_0}{z_1}$ &
    Clip $x$ to the interval $[z_0,z_1]$ \\
    $\lowerclip{x}{z_0}$ &
    $\clip{x}{z_0}{+\infty}$ \\
    $\upperclip{x}{z_1}$ &
    $\clip{x}{-\infty}{z_1}$ \\
    \midrule
    $c$ &
    Battery capacity \\
    $\maxChargeEnergy$ &
    Per-slot maximum chargeable energy \\
    $\maxDischargeEnergy$ & Per-slot maximum dischargeable energy \\
    $r(x)$ &
    Reward function (Eq.~\eqref{eq:reward} and Remark~\ref{re:reward}) \\
    $\mathcal{G}(\sigma)$ &
    Throughput of policy $\sigma$ (Eq.~\eqref{eq:throughput_under_policy}) \\
    $g^*$ &
    Maximum online throughput \\
    \midrule
    $\hat{h}_{q,\hat{\gamma}}(b)$ &
    Relative-value function approximation (Eq.~\eqref{eq:relative_value_function_approximation2}) \\
    $q$ &
    Effectively equivalent linear-policy slope \\
    $\hat{\gamma}$ &
    Effectively equivalent channel SNR coefficient \\
    $\clippedmean(Q,x)$ &
    Dynamic clipped mean of $Q$ with respect to the available charging capacity $x$ (Eq.~\eqref{eq:clipped_mean}) \\
    $\clippedmean(Q,c)$ &
    Clipped mean of $Q$ \\
    DMCR &
    Dynamic mean-to-capacity ratio $\mcratio(Q,x)$ of $Q$ with respect to the available charging capacity $x$ (Eq.~\eqref{eq:dmcr}) \\
    MCR &
    Mean-to-capacity ratio $\mcratio(Q,c)$ \\
    NMCR &
    Nominal mean-to-capacity ratio (Eq.~\eqref{eq:nmcr}) \\
    NSNR &
    Nominal signal-to-noise ratio (Eq.~\eqref{eq:nsnr}) \\
    OCA &
    Optimistic clipped affine policy (Eq.~\eqref{eq:optimistic_clipped_affine_policy}) \\
    RCA &
    Robust clipped affine policy (Eq.~\eqref{eq:robust_clipped_affine_policy}) \\
    \bottomrule
  \end{tabular}
\end{table}

Throughout this paper, the symbol $\geqdef$ denotes a \emph{global} definition, while $\eqdef$ indicates a \emph{local} definition (valid only within a specific scope, such as a section or proof).
Unless specified otherwise, the base of a logarithm is assumed to be Euler's number $\en$ (in upright font).
The probability distributions and the associated notations used in this paper include:
(i) one-point distribution $\onepoint{e}$ at $e \in \real$;
(ii) Bernoulli distribution $\bernoulli_{q,e} \geqdef q\onepoint{e} + (1-q)\onepoint{0}$, $q \in [0,1]$, $e>0$;
(iii) exponential distribution $\expon_\lambda$, with probability density function $f(x) = \lambda\en^{-\lambda x}\indicator\set{x \ge 0}$, $\lambda > 0$;
and (iv) uniform distribution $\uniform_b$ over $[0,b]$.

\section{Problem Formulation}\label{sec:problem_formulation}

\begin{figure}
  \centering
  \includegraphics{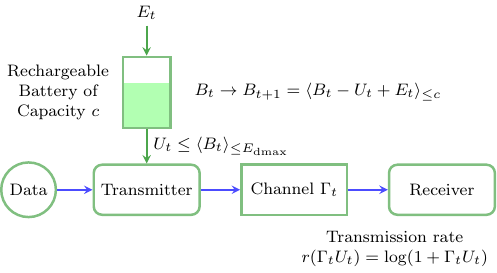}
  \caption{A discrete-time energy harvesting wireless communication system.}\label{fig:system}
\end{figure}

Consider a discrete-time wireless communication system (Fig.~\ref{fig:system}) where a transmitter communicates with a receiver over a fading channel, both with a single antenna.
The transmitter is powered by an energy harvester, which can harvest energy from the environment.
The harvested energy during a time slot is first stored in a rechargeable battery of capacity $c$ and is available for use in the next time slot.
Let $B_t$ and $E_t$ denote the battery level at time $t$ (i.e., the beginning of time slot $t$) and the amount of energy harvested during time slot $t$, respectively.
Then, the battery level at time $t+1$ can be expressed as
\begin{equation}
  B_{t+1}
  = \upperclip*{B_t - \frac{U_t}{\dischargeEfficiency} + \chargeEfficiency \upperclip{E_t}{\maxChargeEnergy}}{c},\label{eq:battery_evolution}
\end{equation}
where $\upperclip{x}{z_1} \geqdef \clip{x}{-\infty}{z_1}$, $U_t$ denotes the amount of energy (actually) consumed by the transmitter during time slot $t$, $\chargeEfficiency \in (0,1]$ denotes the \term{charging efficiency coefficient}, $\dischargeEfficiency \in (0,1]$ denotes the \term{discharging efficiency coefficient}, and $\maxChargeEnergy$ denotes the \term{per-slot maximum chargeable energy}, i.e., the maximum amount of energy that can be charged into the battery during a time slot.
Since $\chargeEfficiency \upperclip{E_t}{\maxChargeEnergy}$ is not available for use in time slot $t$, we must have
\begin{equation}
  U_t/\dischargeEfficiency
  \le B_t,
\end{equation}
the \term{energy-causality} constraint, and
\begin{equation}
  U_t
  \le \maxDischargeEnergy,\label{eq:maximum_dischargeable_energy_constraint}
\end{equation}
the \term{maximum-dischargeable-energy} constraint, where $\maxDischargeEnergy$ denotes the \term{per-slot maximum dischargeable energy}, i.e., the maximum amount of energy that can be discharged from the battery during a time slot.

After simple algebra, we can equivalently write
\iftwocolumn{%
  \begin{multline}
    \upperclip*{B_t - \frac{U_t}{\dischargeEfficiency} + \chargeEfficiency \upperclip{E_t}{\maxChargeEnergy}}{c} \\
    = \frac{1}{\dischargeEfficiency} \upperclip*{\dischargeEfficiency B_t - U_t + \upperclip{\chargeEfficiency \dischargeEfficiency E_t}{\chargeEfficiency \dischargeEfficiency \maxChargeEnergy}}{\dischargeEfficiency c}.
  \end{multline}
}{%
  \begin{equation}
    \upperclip*{B_t - \frac{U_t}{\dischargeEfficiency} + \chargeEfficiency \upperclip{E_t}{\maxChargeEnergy}}{c}
    = \frac{1}{\dischargeEfficiency} \upperclip*{\dischargeEfficiency B_t - U_t + \upperclip{\chargeEfficiency \dischargeEfficiency E_t}{\chargeEfficiency \dischargeEfficiency \maxChargeEnergy}}{\dischargeEfficiency c}.
  \end{equation}
}
Defining $\tilde{B}_t \eqdef \dischargeEfficiency B_t$, $\tilde{U}_t \eqdef U_t$, $\tilde{E}_t \eqdef \upperclip{\chargeEfficiency \dischargeEfficiency E_t}{\chargeEfficiency \dischargeEfficiency \maxChargeEnergy}$, and $\tilde{c} \eqdef \dischargeEfficiency c$ yields
\begin{equation}
  \tilde{B}_{t+1}
  = \upperclip*{\tilde{B}_t - \tilde{U}_t + \tilde{E}_t}{\tilde{c}}.
\end{equation}
Thus, after this change of variables, the battery-evolution model has unit charging/discharging efficiencies;
the transformed per-slot charging limit equals $\chargeEfficiency\dischargeEfficiency\maxChargeEnergy$ and may be absorbed into the (possibly clipped) energy-arrival process.
For analytical convenience, we therefore set $\chargeEfficiency=\dischargeEfficiency=1$ and treat $\maxChargeEnergy$ as effectively infinite, while in simulations, we also use finite $\maxChargeEnergy$ to evaluate the policies' performance under energy arrivals clipped at $\maxChargeEnergy$.

\begin{remark}
  The maximum-dischargeable-energy constraint (Eq.~\eqref{eq:maximum_dischargeable_energy_constraint}) has little effect unless the (effective average) energy-arrival rate is comparable to $\maxDischargeEnergy$.
  If the arrival rate is much smaller than $\maxDischargeEnergy$, the constraint is effectively inactive for an optimal policy;
  if it is much larger, the constraint is binding and the optimal policy is to discharge at the maximum rate $\maxDischargeEnergy$ whenever the battery level exceeds $\maxDischargeEnergy$.
  The only nontrivial case occurs when the energy-arrival rate is close to $\maxDischargeEnergy$.
  In what follows, we first assume that the maximum-dischargeable-energy constraint is inactive, and then discuss how to incorporate it into the derived policies (see Remark~\ref{re:modified_oca_and_rca}).
\end{remark}

Having established the battery evolution and the normalization that sets the charging/discharging efficiencies to one (with any per-slot charging limit absorbed into the energy-arrival process), we turn to the communication model.
Specifically, we assume that the consumed energy $U_t$ is exclusively allocated for data transmission over a Gaussian flat-fading channel.
The channel gain remains constant within each time slot but varies independently and identically across different time slots, corresponding to a slow block-fading scenario.
Then, the instantaneous rate achievable with $U_t$ during time slot $t$ is $r(\Gamma_t U_t)$ nats/s/Hz (i.e., nats per complex channel use) \cite[Eq.~(5.26)]{tse2005fundamentals}, where
\begin{equation}
  r(x)
  \geqdef \log(1 + x),\label{eq:reward}
\end{equation}
$\Gamma_t \geqdef |H_t|^2$ is the channel SNR coefficient at time slot $t$, and $H_t$ is the complex channel gain (with the noise variance normalized to one), assumed to be known at both the transmitter and receiver.
We normalize $\expect \Gamma_t = 1$ under the assumption that $\expect \Gamma_t$ is time-invariant.
There is no loss of generality, as the constant factor can be absorbed into the definitions of all energy quantities (including $U_t$).

\begin{remark}\label{re:reward}
  Although we use the Shannon capacity formula \eqref{eq:reward} for illustration, the methods developed in this work apply to any reward function that is increasing, strictly concave, and continuously differentiable, provided it closely approximates the system's objective.
\end{remark}

Having fixed the reward function in \eqref{eq:reward}, we are now ready to formalize the control problem as an MDP.
At time $t$, the observable state is $S_t\eqdef(B_t,\Gamma_t)$, where $B_t$ is the battery level and $\Gamma_t$ is the channel SNR coefficient.
We assume that the energy-arrival process $(E_t)_{t\ge1}$ and the channel-SNR-coefficient process $(\Gamma_t)_{t\ge1}$ are mutually independent and each is i.i.d.\ over time.
Consequently, the system evolution is Markovian: $S_{t+1}$ depends on $(S_t,U_t)$ only (see Eq.~\eqref{eq:battery_evolution}).

The MDP components are:
\begin{itemize}
  \item State space $\mathcal{S} \geqdef \set{(b, \gamma): b \in [0,c], \gamma \in [0,+\infty)}$.

  \item Action space: the set of all possible energy consumption levels, defined as $\mathcal{U} \geqdef \bigcup_{s \in \mathcal{S}} \mathcal{U}_s = [0,c]$, where $\mathcal{U}_s \geqdef [0,b]$ denotes the set of allowable actions in state $s = (b,\gamma)$.

  \item Transition probability: the probability of moving from state $s = (b,\gamma)$ to a state in $A \in \borel(\mathcal{S})$ after taking action $u \in \mathcal{U}_s$, defined as $p(A|s,u) \geqdef P\set{(\upperclip{b - u + E}{c}, \Gamma) \in A}$, where $\borel(\mathcal{S})$ denotes the Borel $\sigma$-field on $\mathcal{S}$, and $(E,\Gamma) \eqdist (E_t,\Gamma_{t+1})$ (i.e., equal in distribution).

  \item Reward function: the data rate achieved by taking action $u$ in state $s = (b,\gamma)$, namely, $r(\gamma u)$.
\end{itemize}
The goal of the system is to maximize the (long-term expected) throughput
\begin{equation}
  \mathcal{G}((U_t)_{t=1}^\infty)
  \geqdef \liminf_{n \to \infty} \frac{1}{n} \sum_{t=1}^n \expect r(\Gamma_t U_t).
\end{equation}
We focus on stationary (deterministic online)
policies, which are time-invariant and depend only on the current system state.
An admissible stationary policy $\sigma$ is a mapping from $\mathcal{S}$ to $\mathcal{U}$ such that $\sigma(s) \in \mathcal{U}_s$, i.e., $\sigma(b,\gamma) \in [0,b]$.
The collection of all admissible stationary policies is denoted as $\Sigma$.
Then, the throughput under policy $\sigma$ can be expressed as
\begin{equation}
  \mathcal{G}(\sigma)
  \geqdef \liminf_{n \to \infty} \frac{1}{n} \sum_{t=1}^n \expect r(\Gamma_t \sigma(B_t, \Gamma_t)).\label{eq:throughput_under_policy}
\end{equation}
The power control problem is formulated as the following optimization problem.
\begin{problem}
  \label{prob:optimal_policy}
  Find a stationary policy $\sigma$ attaining or approaching $g^* \geqdef \sup_{\sigma \in \Sigma} \mathcal{G}(\sigma)$, the \term{maximum (online) throughput}.
\end{problem}

\section{Clipped Affine Policy}\label{sec:clipped_affine_policy}

In order to solve Problem~\ref{prob:optimal_policy}, we need to solve the Bellman equation given by the next theorem, an easy consequence of~\cite[Thm~6.1]{arapostathis1993discretetime}.

\begin{theorem}
  If there exist a constant $g$ and a bounded function $h: [0,c] \times [0,+\infty) \to \real$ such that
  \begin{equation}
    g + h(b,\gamma)
    = \sup_{u \in [0,b]} (r(\gamma u) + \expect h(\upperclip{b-u+E}{c},\Gamma)), \label{eq:bellman}
  \end{equation}
  where $(E,\Gamma) \eqdist (E_t,\Gamma_{t+1})$, then $g^* = g$.
  Furthermore, if there exists a stationary policy $\sigma$ such that
  \begin{equation}
    g + h(b,\gamma)
    = r(\gamma\sigma(b,\gamma)) + \expect h(\upperclip{b-\sigma(b,\gamma)+E}{c},\Gamma),
  \end{equation}
  then $\mathcal{G}(\sigma) = g^*$.
\end{theorem}

While we can solve the Bellman equation~\eqref{eq:bellman} numerically, finding a closed-form solution to~\eqref{eq:bellman} is challenging, and often impossible.
The main challenge lies in determining $h(b,\gamma)$, known as the \term{relative-value function}.
Note that $h(b, \gamma)$ is not unique, because all pairs $(g, h+\nu)$ with $\nu \in \real$ are solutions to~\eqref{eq:bellman2}.
In this section, we will derive a family of functions such that for any distribution of $(E,\Gamma)$, there exists a function in the family that well approximates (one of) the relative-value functions in the solution to~\eqref{eq:bellman}.

We first consider the case of deterministic channel SNR coefficient, corresponding to the quasi-static fading scenario.
In this case, $\Gamma = \expect \Gamma = 1$ and the Bellman equation~\eqref{eq:bellman} reduces to
\begin{equation}
  g + h(b)
  = \sup_{u \in [0,b]} (r(u) + \expect h(\upperclip{b-u+E}{c})), \label{eq:bellman2}
\end{equation}
where $h(b) \eqdef h(b,1)$ is the reduced form of $h(b,\gamma)$.
The equation~\eqref{eq:bellman2} is still difficult to solve in general, so we focus on two special cases of energy-arrival distributions:
\begin{enumerate}
  \item One-point distribution: $E \sim \onepoint{e}$, where $e \in [0,c]$.
  \item Bernoulli distribution: $E \sim \bernoulli_p$, where $\bernoulli_p \geqdef \bernoulli_{p,c}$ and $p\in (0,1)$.
\end{enumerate}
These cases represent two extremes: the most stable and unstable energy-supply scenarios.
A promising way is to derive a parameterized function that closely approximates the relative-value functions in both scenarios.

For the one-point distribution $\onepoint{e}$, it is clear that the maximum throughput is $r(e)$.
As for the Bernoulli distribution $\bernoulli_p$, the maximum throughput was obtained in~\cite{kazerouni2015optimal,shaviv2015capacity,shaviv2016universally}.
However, their associated relative-value functions have not been investigated.
The next two theorems give the solutions to~\eqref{eq:bellman2} in the two cases, respectively.

\begin{theorem}
  \label{th:onepoint_bellman2_solution}
  For $e \in [0,c]$, we define $g_1 \eqdef r(e)$ and
  \begin{align}
    h_1(x)
    &\eqdef \int_0^x r'(\upperclip{v}{e}) \diff v \label{eq:onepoint_relative_value_function} \\
    &=
    \begin{cases}
      r(x), &\text{$x \in [0,e)$}, \\
      r(e) + r'(e)(x-e), &\text{$x \in [e, +\infty)$}.
    \end{cases}
  \end{align}
  Then, the pair $(g_1, h_1)$ is a solution to~\eqref{eq:bellman2} for $E \sim \onepoint{e}$, and the corresponding optimal policy is
  \begin{equation}
    \clippedGreedyPolicy[e](x)
    \geqdef \upperclip{x}{e},
  \end{equation}
  referred to as the \term{clipped greedy policy}.
  \proofref{Appendix~\ref{app:bellman2_solution}}
\end{theorem}

\begin{theorem}
  \label{th:bernoulli_bellman2_solution}
  For $p\in (0,1)$, we define
  \begin{equation}
    h_2(x)
    \eqdef \sup_{\substack{(u_i)_{i=0}^\infty: u_i\ge 0,\\ \sum_{i=0}^\infty u_i \le x}} \sum_{i=0}^\infty (1-p)^i r(u_i), \label{eq:bernoulli_relative_value_function}
  \end{equation}
  and $g_2 \eqdef ph_2(c)$.
  Then, the pair $(g_2, h_2)$ is a solution to~\eqref{eq:bellman2} for $E \sim \bernoulli_p$, and the corresponding optimal policy is
  \begin{equation}
    \maximinOptimalPolicy[p](x)
    \geqdef \frac{p(x+\tilde{M}(x))}{1-(1-p)^{\tilde{M}(x)}} - 1,\label{eq:maximin_optimal_policy}
  \end{equation}
  referred to as the \term{maximin optimal policy} (\cite[Thm.~1]{yang2020maximin} or~\cite[Cor.~3.28]{yang2025power}), where
  \begin{equation}
    \tilde{M}(x)
    \geqdef \min\set{i \ge 1: [1+p(x+i)](1-p)^i < 1} \ge 1.
  \end{equation}
  Furthermore,
  \begin{equation}
    h_2(x)
    = \sum_{i=0}^{\tilde{M}(x) - 1} (1-p)^i r(\maximinOptimalPolicy[p](\reserve{\maximinOptimalPolicy[p]}^{(i)}(x))),
  \end{equation}
  where $\sigma^{(i)}$ denotes the $i$-th iteration of the function $\sigma$ (with $\sigma^{(0)}(x) \geqdef x$), and
  \begin{equation}
    \reserve{\sigma}(x)
    \geqdef x - \sigma(x).
  \end{equation}
  \proofref{Appendix~\ref{app:bellman2_solution}}
\end{theorem}

Now the problem is to find a parameterized function that well approximates both $h_1$ and $h_2$.
Observe that the shape of $h_1$ and $h_2$ are both determined by their slopes, i.e.,
\begin{equation}
  h_1'(x)
  = r'(\clippedGreedyPolicy[e](x))
  \quad \text{and} \quad
  h_2'(x)
  \eqvar{(a)} r'(\maximinOptimalPolicy[p](x)), \label{eq:relative_value_function_slope}
\end{equation}
respectively, where (a) follows from~\cite[Lem.~3]{yang2020maximin}.
Interestingly, both $\clippedGreedyPolicy[e](x)$ and $\maximinOptimalPolicy[p](x)$ are optimal for one-point and Bernoulli distributions, respectively.

We focus on the Bernoulli case, because $\maximinOptimalPolicy[p](x)$ is proved to be maximin optimal in~\cite[Thm.~1]{yang2020maximin}.
Roughly speaking, it is universally good for any energy-arrival distribution $Q$ with the \term{clipped mean} $\clippedmean(Q,c)=pc$, where
\begin{equation}
  \clippedmean(Q,x)
  \geqdef \expect_{X \sim Q} \upperclip{X}{x}. \label{eq:clipped_mean}
\end{equation}
From~\cite{shaviv2016universally,yang2020maximin}, we know that the maximin optimal policy $\maximinOptimalPolicy[p](x)$ can be well approximated by the \term{fixed fraction policy}
\begin{equation}
  \fixedFractionPolicy[p](x)
  \geqdef px
\end{equation}
for large $x$.
Moreover, \cite{wang2021optimality,garmaroudi2022linear} show that linear policies (including the greedy policy) perform well for general energy-arrival distributions when the slope is properly chosen.
Therefore, replacing $\maximinOptimalPolicy[p](x)$ with a linear policy $qx$ in~\eqref{eq:relative_value_function_slope}, we obtain the following approximate relative-value function:
\begin{equation}
  \hat{h}_q(b)
  \geqdef \int_0^b r'(qx) \diff x
  =
  \begin{dcases}
    \frac{1}{q} r(qb), &\text{$q \in (0,1]$}, \\
    b, &\text{$q = 0$},
  \end{dcases} \label{eq:relative_value_function_approximation}
\end{equation}
where $q$ is termed the \term{effectively equivalent linear-policy slope}.
At first glance, this approximation may not be good for $h_1$.
The next result, however, shows that $\hat{h}_q(b)$ is a solution to~\eqref{eq:bellman2} in a wide sense.
\begin{theorem}
  \label{th:onepoint_bellman2_solution2}
  Suppose that $e \in [0,c]$.
  For $b = b_0 \eqdef e/q \in [e,c]$ with $q \in [e/c,1] \setminus \{0\}$, the pair $(r(e), \hat{h}_q)$ is a solution to~\eqref{eq:bellman2} when $E \sim \onepoint{e}$.
  The corresponding optimal policy is
  \(
  \sigma(x)
  \eqdef qx,
  \)
  which yields the asymptotic behavior
  \begin{equation}
    \lim_{n \to \infty} \phi^{(n)}(x)
    = b_0 \quad \text{for all $x \in [0,c]$}, \label{eq:onepoint_bellman2_solution2.evolution}
  \end{equation}
  where $\phi(x) \eqdef \reserve{\sigma}(x)+e = (1-q)x+e$.
  \proofref{Appendix~\ref{app:bellman2_solution}}
\end{theorem}

Eq.~\eqref{eq:onepoint_bellman2_solution2.evolution} shows that under $\sigma(x) = qx$, the system has a unique fixed point $b_0$, and all other states are transient.
It can be shown by~\cite[Thm.~4.3.4]{hernandez-lerma2003markov} that the system forms an aperiodic positive Harris recurrent Markov chain with the unique invariant probability measure $\onepoint{b_0}$.
Accordingly, the pair $(r(e), \hat{h}_q)$ need only satisfy the Bellman equation~\eqref{eq:bellman2} for $b=b_0$.

Next, we extend~\eqref{eq:relative_value_function_approximation} to the case of random channel SNR coefficient.
Note that the expectation $\expect h(\upperclip{b-u+E}{c},\Gamma)$ in the right-hand side of~\eqref{eq:bellman} can be rewritten as
\begin{equation}
  \expect (\expect (h(\upperclip{b-u+E}{c},\Gamma) | E))
  = \expect \bar{h}(\upperclip{b-u+E}{c}),
\end{equation}
where $\bar{h}(b) \eqdef \expect h(b, \Gamma)$ and the equality follows from the independence between $\Gamma$ and $E$.
We consider the following approximation and upper bound of $\bar{h}(b)$:
\begin{align}
  \bar{h}(b)
  &\approx \expect \left( \frac{1}{q} r(qbf_{\theta}(\Gamma)) \right)
  = \frac{1}{q} \expect r(qbf_{\theta}(\Gamma)) \\
  &\le \frac{1}{q} r(qb\expect(f_{\theta}(\Gamma))) \quad\text{(Jensen's inequality)},\label{eq:relative_value_function_approximation2.pre}
\end{align}
where $f_{\theta}$ is a real-valued function with the parameter $\theta$.
By a second-order Taylor expansion of $r(x)$ around $\expect X$, the Jensen gap $r(qb\expect X)-\expect r(qbX)$ is governed (to leading order) by the product $\lvert r''(qb\expect X)\rvert\,\mathrm{Var}(X)$ with $X=f_{\theta}(\Gamma)$.
Hence, the upper bound~\eqref{eq:relative_value_function_approximation2.pre} is relatively tight when $f_{\theta}(\Gamma)$ exhibits low variability or when $r$ is nearly linear over the relevant range of $qbX$.
Furthermore, the approximation can be improved by introducing an arbitrary effective constant $\hat{\gamma}$ and an additive offset $\nu$, yielding
\begin{equation}
  \expect \left( \frac{1}{q} r(qbf_{\theta}(\Gamma)) \right)
  \approx \frac{1}{q} r(\hat{\gamma}qb) + \nu.
\end{equation}
Since the offset $\nu$ is immaterial in the Bellman equation, we obtain the following fundamental approximation of $\expect h(b,\Gamma)$:
\begin{equation}
  \hat{h}_{q,\hat{\gamma}}(b)
  \geqdef
  \begin{dcases}
    \frac{1}{q} r(\hat{\gamma}qb), &\text{$q \in (0,1]$}, \\
    \lim_{q \downarrow 0} \hat{h}_{q,\hat{\gamma}}(b) = \hat{\gamma} b, &\text{$q = 0$},
  \end{dcases} \label{eq:relative_value_function_approximation2}
\end{equation}
where $\hat{\gamma}$ is termed the \term{effectively equivalent channel SNR coefficient}.

Based on the approximation~\eqref{eq:relative_value_function_approximation2}, the right-hand side of~\eqref{eq:bellman} can be reformulated approximately as the following optimization problem.

\begin{problem}
  \label{prob:approximate_optimal_action}
  For $b \in [0,c]$,
  \begin{align}
    \text{maximize} \quad &r(\gamma u) + \expect \hat{h}_{q,\hat{\gamma}}(\upperclip{b-u+E}{c}), \\
    \text{subject to} \quad &u \in [0,b], \notag
  \end{align}
  where $E \eqdist E_t$.
\end{problem}

For a general distribution of $E$, Problem~\ref{prob:approximate_optimal_action} is difficult to solve.
We thus turn to maximizing its lower or upper bound.
Since $\hat{h}_{q,\hat{\gamma}}(b)$ is a concave function of $b$, we can use Jensen's inequality to obtain the following upper bound:
\begin{align}
  \expect \hat{h}_{q,\hat{\gamma}}(\upperclip{b-u+E}{c})
  &\le \hat{h}_{q,\hat{\gamma}}(\expect \upperclip{b-u+E}{c}), \notag \\
  &= \hat{h}_{q,\hat{\gamma}}(b-u+\expect \upperclip{E}{c-b+u}).
\end{align}
On the other hand, since
\begin{equation}
  f(x)
  \eqdef \frac{c-x}{c-b+u} \hat{h}_{q,\hat{\gamma}}(b-u) + \frac{x-b+u}{c-b+u} \hat{h}_{q,\hat{\gamma}}(c)
\end{equation}
is the lower convex envelope of $\hat{h}_{q,\hat{\gamma}}(x)$ on $[b-u,c]$, we can also use Jensen's inequality to obtain the following lower bound (see also~\cite[Prop.~3.3]{yang2025power} for Jensen's inequality for arbitrary functions):
\begin{align}
  \expect \hat{h}_{q,\hat{\gamma}}(\upperclip{b-u+E}{c})
  &\ge \expect f(\upperclip{b-u+E}{c}) \notag \\
  &\ge f(\expect \upperclip{b-u+E}{c}) \notag \\
  &= f(b-u+\expect \upperclip{E}{c-b+u}) \notag \\
  &= \left(1-\frac{\expect \upperclip{E}{c-b+u}}{c-b+u}\right) \hat{h}_{q,\hat{\gamma}}(b-u) \notag \\
  &\quad + \frac{\expect \upperclip{E}{c-b+u}}{c-b+u} \hat{h}_{q,\hat{\gamma}}(c).
\end{align}
Problem~\ref{prob:approximate_optimal_action} can then be approximated by one of two formulations: an \emph{optimistic} formulation based on a certainty-equivalence-type approximation, and a \emph{pessimistic} formulation based on worst-case analysis.

\begin{problem}[Optimistic Formulation]
  \label{prob:optimistic_optimal_action}
  For $b \in [0,c]$,
  \begin{align}
    \text{maximize} \quad &r(\gamma u) + \hat{h}_{q,\hat{\gamma}}(b-u+\clippedmean(P_E,c-b+u)), \\
    \text{subject to} \quad &u \in [0,b], \notag
  \end{align}
  where $P_E$ denotes the distribution of $E$, and $\clippedmean(Q,x)$ is defined by~\eqref{eq:clipped_mean} and gives the \term{dynamic clipped mean} of $Q$ with respect to the available charging capacity $x$.
\end{problem}

\begin{problem}[Pessimistic Formulation]
  \label{prob:pessimistic_optimal_action}
  For $b \in [0,c]$,
  \begin{align}
    \text{maximize} \quad &r(\gamma u) + (1-\mcratio(P_E,c-b+u)) \hat{h}_{q,\hat{\gamma}}(b-u) \notag \\
    &+ \mcratio(P_E,c-b+u)\hat{h}_{q,\hat{\gamma}}(c), \\
    \text{subject to} \quad &u \in [0,b], \notag
  \end{align}
  where
  \begin{equation}
    \mcratio(Q,x)
    \geqdef
    \begin{dcases}
      \frac{\clippedmean(Q,x)}{x}, &\text{$x>0$}, \\
      \lim_{x\downarrow 0} \mcratio(Q,x) = 1-Q(\{0\}), &\text{$x=0$},
    \end{dcases}\label{eq:dmcr}
  \end{equation}
  coined the \term{dynamic mean-to-capacity ratio} (DMCR) of $Q$ with respect to the available charging capacity $x$, a generalization of the mean-to-capacity ratio (MCR) $\mcratio(Q,c)$ in~\cite[Def.~3]{yang2020maximin} or~\cite[Def.~2.3]{yang2025power}.
\end{problem}

Note that
\begin{equation}
  \clippedmean(\delta_e,x)
  = \upperclip{e}{x}
  \quad \text{and} \quad
  \mcratio(\bernoulli_p,x)
  = p,
\end{equation}
which provide the typical cases of the optimistic and pessimistic formulations, respectively.
Corresponding to these two cases, the next two theorems provide the optimal solutions to Problems~\ref{prob:optimistic_optimal_action} and~\ref{prob:pessimistic_optimal_action}, respectively.

\begin{theorem}[Optimistic Clipped Affine Policy]
  \label{th:optimistic_clipped_affine_policy}
  If $\clippedmean(P_E,x) = \upperclip{e}{x}$ with $e \in [0,c]$, then the optimal action for Problem~\ref{prob:optimistic_optimal_action} is
  \begin{equation}
    \optimisticClippedAffinePolicy[e,q,\hat{\gamma}](b,\gamma)
    \geqdef
    \begin{dcases}
      \clip*{\frac{q(b+e)-1/\gamma+1/\hat{\gamma}}{1+q}}{b_0(e)}{b}, &\text{$\gamma > 0$}, \\
      \lim_{\gamma \downarrow 0} \optimisticClippedAffinePolicy[e,q,\hat{\gamma}](b,\gamma) = b_0(e), &\text{$\gamma = 0$},
    \end{dcases}\label{eq:optimistic_clipped_affine_policy}
  \end{equation}
  where $\optimisticClippedAffinePolicy$ is coined the \term{optimistic clipped affine (OCA) policy},
  \begin{equation}
    b_0(e)
    \eqdef \lowerclip{b+e-c}{0},
  \end{equation}
  and $\lowerclip{x}{z_0} \geqdef \clip{x}{z_0}{+\infty}$.
  \proofref{Appendix~\ref{app:clipped_affine_policy}}
\end{theorem}

\begin{theorem}[Robust Clipped Affine Policy]
  \label{th:robust_clipped_affine_policy}
  If $\mcratio(P_E,x) = p \in [0,1)$ for $x \le c$, then the optimal action for Problem~\ref{prob:pessimistic_optimal_action} is
  \begin{equation}
    \robustClippedAffinePolicy[p,q,\hat{\gamma}](b,\gamma)
    \geqdef
    \begin{dcases}
      \clip*{\frac{qb-(1-p)/\gamma+1/\hat{\gamma}}{1-p+q}}{0}{b}, &\text{$\gamma > 0$}, \\
      \lim_{\gamma \downarrow 0} \robustClippedAffinePolicy[p,q,\hat{\gamma}](b,\gamma) = 0, &\text{$\gamma = 0$},
    \end{dcases}\label{eq:robust_clipped_affine_policy}
  \end{equation}
  where $\robustClippedAffinePolicy$ is coined the \term{robust clipped affine (RCA) policy}.
  \proofref{Appendix~\ref{app:clipped_affine_policy}}
\end{theorem}

\begin{remark}[OCA and RCA policies with a per-slot maximum dischargeable energy]\label{re:modified_oca_and_rca}
  In the proofs of Theorems~\ref{th:optimistic_clipped_affine_policy} and~\ref{th:robust_clipped_affine_policy}, the optimal actions are obtained by maximizing a concave objective under the constraint $u\in[0,b]$.
  If a per-slot maximum dischargeable energy $\maxDischargeEnergy$ is imposed, then the feasible set $[0,b]$ becomes $[0,\min\{b,\maxDischargeEnergy\}]$, and the maximizer over this restricted feasible set is obtained by clipping the OCA/RCA policy output to the interval $[0,\maxDischargeEnergy]$, namely,
  \begin{equation}
    \upperclip*{\optimisticClippedAffinePolicy[e,q,\hat{\gamma}](b,\gamma)}{\maxDischargeEnergy}
    \quad \text{and} \quad
    \upperclip*{\robustClippedAffinePolicy[p,q,\hat{\gamma}](b,\gamma)}{\maxDischargeEnergy}.
  \end{equation}
\end{remark}

As shown in subsequent sections, the OCA and RCA policies with appropriately chosen parameters can achieve near-optimal performance for various energy-arrival distributions.
For now, we illustrate some simple but fundamental policies using basic parameter choices.

For the optimistic clipped affine policy, we can set $q = 1$ and $\hat{\gamma} = 1$.
Then, we have
\begin{equation}
  \optimisticClippedAffinePolicy[e,q,1](b,\gamma)
  = \clip*{\frac{b+e-1/\gamma+1}{2}}{b_e}{b}.
\end{equation}
When $\gamma = 1$, the policy reduces to
\begin{equation}
  \optimisticClippedAffinePolicy[e,1,1](b,1)
  = \clip*{\frac{b+e}{2}}{b_e}{b}
  =
  \begin{dcases}
    b, &\text{$b \le e$}, \\
    \frac{b+e}{2}, &\text{$b > e$},
  \end{dcases}
\end{equation}
recovering the optimal offline policy with perfect knowledge of the current battery level $b$ and the next energy arrival $e$ in the quasi-static fading scenario (see~\cite{tutuncuoglu2012optimum,yang2025power}).

For the robust clipped affine policy, we can set $q = p$ and $\hat{\gamma} = 1$.
Then, we have
\begin{equation}
  \robustClippedAffinePolicy[p,p,1](b,\gamma)
  = \clip*{pb-(1-p)/\gamma+1}{0}{b}.
\end{equation}
When $\gamma = 1$, the policy reduces to
\begin{align}
  \robustClippedAffinePolicy[p,p,1](b,1)
  &= \upperclip*{p(b+1)}{b} \notag \\
  &=
  \begin{cases}
    b, &\text{$b \le p/(1-p)$}, \\
    p(b+1), &\text{$b > p/(1-p)$},
  \end{cases}
\end{align}
which recovers the \term{two-piece fixed fraction policy}~\cite[Sec.~3.4.4]{yang2025power}, a hybrid of the greedy and fixed fraction policies that outperforms both.

\section{Practical Power Control: Closed‑form Policies and Adaptive Online Schemes}\label{sec:practical_power_control}

\subsection{Closed‑form Clipped‑Affine Policies Based on the Maximin Optimal Linear Policy}\label{sec:closed_form_policies}

The parameters of the OCA and RCA policies fall into two groups:
distribution-dependent parameters that reflect the energy-arrival statistics ($e$ for OCA or $p$ for RCA), and parameters used for relative-value fitting ($q$ and $\hat{\gamma}$).

Although in Theorems~\ref{th:optimistic_clipped_affine_policy} and~\ref{th:robust_clipped_affine_policy}, the parameters $e$ and $p$ are defined respectively as the constant dynamic clipped mean and the constant DMCR of certain special energy-arrival distributions, they are intended to serve as fixed estimates of dynamic clipped mean and DMCR for a general energy-arrival distribution.
When the detailed knowledge of an energy-arrival distribution is unavailable, a reasonable choice is to take $e$ as the clipped mean and $p$ as the MCR of the distribution.

For the effectively equivalent channel SNR coefficient $\hat{\gamma}$, a natural choice is the expected channel SNR coefficient, which captures the average channel quality.
The selection of the effectively equivalent linear-policy slope $q$ is more delicate.
Intuitively, $q$ should closely approximate the effectively equivalent linear slope of the (unknown) optimal policy.
A natural choice is the slope of the \term{maximin optimal linear policy} proposed in~\cite{garmaroudi2022linear}.
That linear policy achieves performance close to the maximin optimal policy.
Its slope is defined by
\begin{equation}
  \maximinOptimalLinear(c,p)
  \geqdef \arg\max_{s\in [0,1]} \underline{\Gamma}(c,p,s),\label{eq:mol_definition}
\end{equation}
where $p$ denotes the MCR of the energy-arrival distribution, and
\begin{equation}
  \underline{\Gamma}(c,p,s)
  \geqdef \sum_{i=0}^\infty p(1-p)^{i} r(cs(1-s)^{i})\label{eq:lower_gamma}.
\end{equation}
For $c\leq p/(1-p)$, $\maximinOptimalLinear(c,p)=1$.
For $c > p/(1-p)$, no closed-form expression is known for $\maximinOptimalLinear(c,p)$.
A practical approach is to precompute $\maximinOptimalLinear(c,p)$ on a grid, store the values in a lookup table, and use interpolation at runtime to obtain intermediate values.
Alternatively, we can use the following approximation:
\begin{equation}
  \maximinOptimalLinearHat(c,p)
  \geqdef \upperclip*{\frac{\maximinOptimalPolicy[p](c)}{c} + 0.043}{1},\label{eq:mol_approximation}
\end{equation}
where $\maximinOptimalPolicy[p]$ is the maximin optimal policy in~\eqref{eq:maximin_optimal_policy}.

Based on the above discussion, we construct two closed‑form clipped‑affine policies, $\ocaOl[Q]$ and $\rcaOl[Q]$, by instantiating the OCA and RCA templates with the recommended parameter choices.
They are summarized in Table~\ref{tab:closed_form_policies} and are referred to as OCA‑OL and RCA‑OL, where ``OL'' denotes the maximin optimal linear policy.

\begin{table}
  \centering
  \caption{Closed‑Form Clipped‑Affine Policies for Energy-Arrival Distribution $Q$}
  \label{tab:closed_form_policies}
  \tablelinespread
  \begin{tabular}{cccc}
    \toprule
    Policy & $e$ or $p$ & $q$ & $\hat{\gamma}$ \\
    \midrule
    $\ocaOl[Q] \geqdef \optimisticClippedAffinePolicy[e,q,\hat{\gamma}]$ &
    $e=\clippedmean(Q,c)$ &
    $\maximinOptimalLinear(c,\mcratio(Q,c))$ &
    $\expect\Gamma$ \\
    $\rcaOl[Q] \geqdef \robustClippedAffinePolicy[p,q,\hat{\gamma}]$ &
    $p=\mcratio(Q,c)$ &
    $\maximinOptimalLinear(c,\mcratio(Q,c))$ &
    $\expect\Gamma$ \\
    \bottomrule
  \end{tabular}
\end{table}

\subsection{Adaptive Clipped‑Affine Policies: Online Parameter Tuning}

In this subsection, we present adaptive schemes that tune the parameters of the OCA and RCA policies online, using observed data to estimate the energy-arrival statistics and to learn the effectively equivalent linear-policy slope and channel SNR coefficient.

It is clear that using clipped mean and MCR to estimate the dynamic clipped mean and MCR may not be very accurate, because the available charging capacity
\begin{equation}
  C_t
  \geqdef c-B_t+U_t
\end{equation}
is random and may be much smaller than the battery capacity $c$.
As a result, the clipped mean may overestimate the dynamic clipped mean, and the MCR may underestimate the DMCR.
An obvious solution to this issue is to estimate the dynamic clipped mean and DMCR directly from the observed battery evolution and available charging capacity.

For a general energy-arrival distribution $Q$, the parameter $e$ can be estimated by the dynamic clipped mean of $Q$ with respect to high available charging capacity, i.e.,
\begin{align}
  e
  &\approx \expect (\clippedmean(Q, C_t) | C_t \ge \expect C_t) \notag \\
  &= \expect (\upperclip{E_t}{C_t} | C_t \ge \expect C_t) \notag \\
  &= \expect (B_{t+1}-B_t+U_t | C_t \ge \expect C_t). \label{eq:estimation_of_e}
\end{align}
The parameter $p$ can be estimated by the expected DMCR of $Q$ with respect to the random available charging capacity, i.e.,
\begin{align}
  p
  &\approx \expect \mcratio(Q, C_t)
  = \expect \left( \frac{\upperclip{E_t}{C_t}}{C_t} \right) \notag \\
  &= \expect \left( \frac{B_{t+1}-B_t+U_t}{C_t} \right). \label{eq:estimation_of_p}
\end{align}

By maintaining running-average counterparts of~\eqref{eq:estimation_of_e} and~\eqref{eq:estimation_of_p}, we obtain online estimates of $e$ and $p$ at each time slot.
Injecting these estimators into the OCA-OL and RCA-OL rules yields their adaptive variants, denoted OCA-OL-SA and RCA-OL-SA, as summarized in Algorithm~\ref{alg:ca-ol-a}, where ``SA'' denotes semi-adaptive.

\begin{algorithm}[H]
  \caption{OCA-OL-SA and RCA-OL-SA: Semi-Adaptive OCA and RCA Policies Based on Maximin Optimal Linear-Policy Slope}\label{alg:ca-ol-a}
  \begin{algorithmic}[1]
    \State Hyperparameters: learning rate $\eta > 0$
    \State Initialization: $q \gets \maximinOptimalLinear(c,\mcratio(Q,c))$, $\hat{\gamma} \gets \expect \Gamma$
    \State Policy-specific Initialization:
    \begin{align*}
      &e \gets 0,\ \hat{c} \gets 0 &\text{for OCA policy} \\
      &p \gets 0 &\text{for RCA policy}
    \end{align*}
    \State Observe the initial state $(B,\Gamma)$
    \For{each step}
      \State Take action
      \[
        U
        \eqdef
        \begin{cases}
          \optimisticClippedAffinePolicy[e,q,\hat{\gamma}](B,\Gamma) &\text{for OCA policy}\\
          \robustClippedAffinePolicy[p,q,\hat{\gamma}](B,\Gamma) &\text{for RCA policy}
        \end{cases}
      \]
      \State Observe the next state $(B',\Gamma')$
      \State $E \gets B' - B + U$, $C \gets c - B + U$
      \State \algIndentParagraph{Update $e$ or $p$:}
      \begin{align*}
        &\left.
           \begin{aligned}
             &e \gets e + \eta (E - e) \indicator\set{C \ge \hat{c}} \\
             &\hat{c} \gets \hat{c} + \eta (C - \hat{c})
           \end{aligned}\right\}
        &\text{for OCA policy} \\
        &\left.p \gets p + \eta (E/C - p)\right. &\text{for RCA policy}
      \end{align*}
      \State $(B,\Gamma) \gets (B',\Gamma')$
    \EndFor
  \end{algorithmic}
\end{algorithm}

The simulation results indicate that the adaptive estimate of $p$ markedly improves the performance of RCA‑OL under non‑Bernoulli energy‑arrival distributions, whereas the adaptive estimate of $e$ yields only marginal gains for OCA‑OL.
Taken together, these results show that RCA‑based policies outperform OCA‑based policies across both closed‑form and adaptive variants; accordingly, we restrict subsequent attention to improving parameter tuning for RCA policies.
This suggests that, as in the quasi-static fading case, the policy derived from worst-case analysis remains near-optimal across a broad range of energy-arrival distributions;
hence, worst-case analysis may serve as a fundamental tool for designing robust and effective policies in battery-limited energy-harvesting communication systems.

The RCA-OL-SA policy is not fully adaptive because the parameters $q$ and $\hat{\gamma}$ remain fixed at their initial values.
A simple improvement is to replace the expected channel SNR coefficient $\expect \Gamma$ with a running-average estimate; the more challenging part is adapting $q$.
A simple approach is to update $q$ periodically using the estimated DMCR $p$ instead of the MCR.
However, computing the maximin‑optimal linear slope $\maximinOptimalLinear(c,p)$ at every update can be computationally intensive.
To reduce this computational burden, we instead use the approximation $\maximinOptimalLinearHat(c,p)$.
This yields a fully adaptive variant of RCA‑OL, denoted RCA‑OLA‑A (OLA: maximin optimal linear-policy-slope approximation; A: adaptive), as summarized in Algorithm~\ref{alg:rca-ola-a}.

\begin{algorithm}[H]
  \caption{RCA-OLA-A: Adaptive RCA Policy Based on Maximin Optimal Linear-Policy-Slope Approximation}\label{alg:rca-ola-a}
  \begin{algorithmic}[1]
    \State Hyperparameters: learning rate $\eta > 0$, update interval $T > 0$ for $q$
    \State Initialization: $p \gets 0$, $q \gets 1$, $\hat{\gamma} \gets \hat{\gamma}_0$ (e.g., $\hat{\gamma}_0 = \expect \Gamma$)
    \State Observe the initial state $(B,\Gamma)$
    \For{each step}
      \State Take action $U\eqdef \robustClippedAffinePolicy[p,q,\hat{\gamma}](B,\Gamma)$
      \State Observe the next state $(B',\Gamma')$
      \State $E \gets B' - B + U$, $C \gets c - B + U$
      \State $p \gets p + \eta (E/C - p)$
      \State $\hat{\gamma} \gets \hat{\gamma} + \eta (\Gamma - \hat{\gamma})$
      \State Every $T$ steps, set $q \gets \maximinOptimalLinearHat(c,p)$
      \State $(B,\Gamma) \gets (B',\Gamma')$
    \EndFor
  \end{algorithmic}
\end{algorithm}

Although RCA-OLA-A performs very well in simulations, it still relies on the assumption that the maximin-optimal linear slope and the expected channel-SNR coefficient are good approximations for the effectively equivalent linear-policy slope and channel-SNR coefficient, respectively.
This assumption may not hold in general, and hence the performance of RCA-OLA-A may degrade in certain scenarios.
To design a more flexible and powerful data-driven adaptive scheme, rather than relying on closed-form expressions such as~\eqref{eq:mol_definition}, we leverage the average-reward RL framework~\cite[Sec.~10.3]{sutton2018reinforcement} to tune the parameters $q$ and $\hat{\gamma}$ of the RCA policy.

\begin{figure}[htbp]
  \centering
  \includegraphics{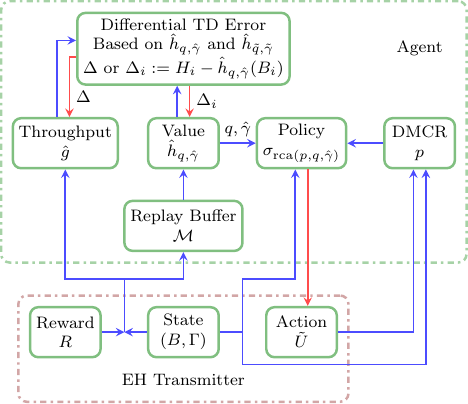}
  \caption{An illustration of Algorithm~\ref{alg:rca-rl}.}\label{fig:rl}
\end{figure}

Starting from RCA-OL-SA, we employ the standard relative-value estimation framework for average-reward RL and update $q$ and $\hat{\gamma}$ online by minimizing the mean-squared differential temporal-difference (TD) error, thereby yielding an adaptive RCA policy, denoted RCA-RL, as summarized in Algorithm~\ref{alg:rca-rl} and Fig.~\ref{fig:rl}.
A version of this algorithm was previously disclosed in~\cite{yang2025power.patent}.

\begin{algorithm}[H]
  \caption{RCA-RL: Adaptive RCA Policy Based on Reinforcement Learning}\label{alg:rca-rl}
  \begin{algorithmic}[1]
    \State Hyperparameters: learning rates $\eta_1, \eta_2, \eta_3 > 0$, replay buffer capacity $M \ge 1$, minibatch size $N \ge 1$, exploration probability $\epsilon \in [0,1)$, warm-up period $D \ge 1$, and target-network update interval $T_{\text{target}} > 0$ (for target relative-value function)
    \State Initialize DMCR estimate $p\in [0,1]$ (e.g., $p \gets 0$)
    \State Initialize throughput estimate $\hat{g} \ge 0$ (e.g., $\hat{g} \gets 0$)
    \State Initialize replay buffer $\mathcal{M}$ as an empty first-in-first-out (FIFO) queue with capacity $M$
    \State Initialize online relative-value function $\hat{h}_{q,\hat{\gamma}}(b)$ with parameters, e.g., $(q,\hat{\gamma}) \gets (0.5,\expect \Gamma)$
    \State Initialize target relative-value function $\hat{h}_{\tilde{q},\tilde{\gamma}}(b)$ with parameters, e.g., $(\tilde{q},\tilde{\gamma}) \gets (1,\expect \Gamma)$
    \State Observe the initial state $(B,\Gamma)$
    \For{$t=1,2,3,\ldots$}
      \State Take action $\tilde{U} \sim (1-\epsilon)
      \onepoint{U} + \epsilon \uniform_B$, where
      \[
        U \eqdef \robustClippedAffinePolicy[p,q,\hat{\gamma}](B,\Gamma)\label{line:action_selection}
      \]
      \State \algIndentParagraph{Observe the reward $R$ (modeled as $r(\Gamma U)$) and the next state $(B',\Gamma')$}
      \State \algIndentParagraph{Push $(B,R,B')$ into $\mathcal{M}$ and pop its oldest entry if the buffer size exceeds $M$}
      \State $E \gets B' - B + \tilde{U}$, $C \gets c - B + \tilde{U}$
      \State $p \gets p + \eta_1 (E/C - p)$
      \State $\Delta \gets R - \hat{g} + \hat{h}_{\tilde{q},\tilde{\gamma}}(B') - \hat{h}_{\tilde{q},\tilde{\gamma}}(B)$
      \State $\hat{g} \gets \hat{g} + \eta_2 \Delta$
      \If{$t = D$}
        \State $q \gets p$
      \ElsIf
          {$t > D$}
        \State \algIndentParagraph[3em]{Sample random minibatch of $N$ entries $(B_i, R_i, B'_i)$ from $\mathcal{M}$}
        \State $H_i \gets R_i - \hat{g} + \hat{h}_{\tilde{q},\tilde{\gamma}}(B_i'), \quad i=1,\ldots,N$
        \State Perform a gradient descent step on
        \[
          L(q,\hat{\gamma})
          \eqdef \frac{1}{N\hat{g}^2} \sum_{i=1}^N \left(H_i-\hat{h}_{q,\hat{\gamma}}(B_i)\right)^2
        \]
        \algIndent\algIndent%
        with the learning rate $\eta_3$
        \If{$(t - D) \bmod T_{\text{target}} = 0$}
          \State $(\tilde{q}, \tilde{\gamma}) \gets (q,\hat{\gamma})$
        \EndIf
      \EndIf
      \State $(B,\Gamma) \gets (B',\Gamma')$
    \EndFor
  \end{algorithmic}
\end{algorithm}

Most components of Algorithm~\ref{alg:rca-rl} follow standard average-reward RL practice, including $\epsilon$-greedy exploration, (experience) replay buffer, minibatch updates, and the use of a target relative-value function; see, e.g.,~\cite[Sec.~10.3 and Sec.~16.5]{sutton2018reinforcement} and the standard deep RL literature for related stabilization techniques.
We therefore focus on the design choices that are specific to our problem.
\begin{enumerate}
  \item A distinctive design feature and computational advantage of Algorithm~\ref{alg:rca-rl} is that, by using the worst-case Bellman maximizer (RCA policy), it avoids gradient-based policy updates typical of policy-gradient methods (e.g., deep deterministic policy gradient, DDPG) and avoids action selection via $\arg\max$ over a learned action-value function typical of value-based methods (e.g., deep Q-network, DQN).
  To some extent, the RL component is used only to estimate the policy parameters $q$ and $\hat{\gamma}$ from interaction data, i.e., to fit the relative-value approximation in~\eqref{eq:relative_value_function_approximation2}.
  The complexity for fitting only two parameters is much lower than that for fitting a general function approximator in most RL algorithms.
  Moreover, although we use minibatch updates with replay buffer, our simulations indicate that a simple online update with minibatch size one and a small replay buffer also performs well in practice, provided that adaptive gradient methods (e.g., Adam~\cite{kingma2017adam}) are used to stabilize the updates.

  \item We introduce a warm-up period $D$ so that the online estimates $p$ and $\hat{g}$ can accumulate sufficient experience before they are used to update the parameters $(q,\hat{\gamma})$ of the function $\hat{h}_{q,\hat{\gamma}}$.
  This warm-up period helps avoid premature policy updates caused by noisy early estimates.

  \item When $t=D$, we set $q \gets p$, which provides a natural warm start by aligning the initial slope parameter with the current DMCR estimate.

  \item The loss $L(q,\hat{\gamma})$ (line~22 of Algorithm~\ref{alg:rca-rl}) is normalized by $\hat{g}^2$ to reduce sensitivity to the throughput scale and improve the conditioning of the gradient update.
  This choice is motivated by the fact that the differential TD error $H_i-\hat{h}_{q,\hat{\gamma}}(B_i)$ scales with the relative value over the random battery state, which is typically of the same order as the average reward (throughput) $g$:
  \begin{equation}
    g
    = \expect h(\upperclip{E}{c}, \Gamma) - \expect h(0, \Gamma)
    \approx \expect \hat{h}_{q,\hat{\gamma}}(\upperclip{E}{c}),
  \end{equation}
  as implied by the Bellman equation~\eqref{eq:bellman} and the relative-value approximation in~\eqref{eq:relative_value_function_approximation2}.
  This normalization is helpful for stochastic gradient descent (SGD) with a fixed learning rate, but it may be unnecessary for adaptive gradient methods with second-moment moving-average normalization, such as Adam~\cite{kingma2017adam}, which already adapt the effective step size to the gradient scale.
\end{enumerate}

\begin{remark}
  Although Algorithm~\ref{alg:rca-rl} employs $\epsilon$-greedy exploration, our simulation results suggest that exploration is unnecessary in practice, as the algorithm still performs well even when exploration is disabled.
\end{remark}

\begin{remark}
  Note that the parameters $q$ and $\hat{\gamma}$ in Algorithm~\ref{alg:rca-rl} have constrained ranges: $q \in [0,1]$ and $\hat{\gamma} > 0$.
  Special care must be taken during optimization to ensure updates respect these constraints.
  We adopt reparameterization in the simulation code, transforming $q$ and $\hat{\gamma}$ via unconstrained proxy variables $(\theta_1, \theta_2) \in \real^2$:
  \begin{align}
    q
    &= \sigmoid(\theta_1) \in (0,1), \\
    \hat{\gamma}
    &= \softplus(\theta_2) \in (0,+\infty),
  \end{align}
  where
  \begin{align}
    \sigmoid(\theta)
    &\geqdef \frac{1}{1+\en^{-\theta}}, \\
    \softplus(\theta)
    &\geqdef \log(1+\en^{\theta}).
  \end{align}
  Similar techniques will be tacitly used in the sequel for all parameters with constrained ranges.
\end{remark}

To highlight the advantages of RCA-RL, we conclude this section by comparing it with representative deep RL algorithms in Table~\ref{tab:rl_overhead_comparison}.
A simulation-based performance comparison is presented in Sec.~\ref{subsec:rca_rl_vs_generic_rl}.

\begin{table*}[t]
  \centering
  \caption{Comparison of representative deep RL methods and the proposed RCA-RL scheme.}
  \label{tab:rl_overhead_comparison}
  \tablelinespread
  \begin{tabular}{p{2.5cm} p{4.5cm} p{4.1cm} p{4.5cm}}
    \toprule
    RL Method & Value Evaluation & Policy Improvement & Data Efficiency \\
    \midrule
    Value-based RL\newline (e.g., DQN) &
    High-dimensional: Learns an action-value function $Q(s,a)$ using deep neural networks (DNNs). &
    Exhaustive maximization: Requires $\arg\max_a Q(s,a)$, which can be expensive in large discrete or continuous action spaces. &
    Low: Requires large replay buffers and moderate minibatches to stabilize training. \\
    Actor-critic\newline (e.g., Proximal Policy Optimization (PPO), DDPG) &
    High-dimensional: Estimates a state-value function $V(s)$ or action-value function $Q(s,a)$ via DNNs. &
    Gradient-based: high-dimensional search using stochastic gradient ascent. &
    Low: Relies on large minibatches and large replay buffers to stabilize high-dimensional gradient updates. \\
    \textbf{RCA-RL} &
    \textbf{Low-dimensional}: Learns only two scalar parameters $(q,\hat{\gamma})$ for the relative-value approximation $\hat{h}_{q,\hat{\gamma}}$. &
    \textbf{Analytical}: Closed-form worst-case maximizer (RCA) with performance improved via better estimates of the DMCR $p$ and the parameters $(q,\hat{\gamma})$ (from value evaluation). &
    \textbf{Exceptional}: Works well with \textbf{minibatch size one} and a samll replay buffer capacity (e.g., 16); requires few samples for learning. \\
    \bottomrule
  \end{tabular}
\end{table*}

\section{Prediction-Based Extensions for Scenarios with Contextual Information}\label{sec:prediction_based_extensions}

Compared with real-world scenarios, the MDP model in Sec.~\ref{sec:problem_formulation} may be overly idealized, as it assumes future energy arrivals and channel states are independent of all causally available information.
Practical EH communication systems often possess partial knowledge of future energy arrivals or channel states.
To exploit this knowledge for performance gains, we incorporate it into the model by upgrading the Bellman equation~\eqref{eq:bellman} and its approximate solution (RCA policy) to their contextual counterparts.
In particular, we need to find the conditional expectations of all involved parameters given the contextual information.
Guided by the physical interpretation of the RCA-policy parameters (Table~\ref{tab:physical_interpretation}), this design process naturally leads to a prediction-based power-control approach.
We illustrate this approach with four representative examples that extend the RCA-RL scheme.
The core ideas in these examples can be applied to other schemes and scenarios.

\begin{table}
  \centering
  \caption{Physical Interpretations of Clipped-Affine-Policy Parameters}\label{tab:physical_interpretation}
  \tablelinespread
  \begin{tabular}{cp{1.7cm}p{4.5cm}}
    \toprule
    Parameter &
    Definition &
    Physical Interpretation \\
    \midrule
    $p$ &
    Thm.~\ref{th:robust_clipped_affine_policy} &
    DMCR of the energy-arrival distribution (in the current time slot) \\
    $q$ &
    Eq.~\eqref{eq:relative_value_function_approximation2},\par Probs.~\ref{prob:optimistic_optimal_action} and~\ref{prob:pessimistic_optimal_action} &
    Approximate fraction of energy consumption (in the next time slot) under the optimal policy \\
    $\hat{\gamma}$ &
    Eq.~\eqref{eq:relative_value_function_approximation2},\par Probs.~\ref{prob:optimistic_optimal_action} and~\ref{prob:pessimistic_optimal_action} &
    Effectively equivalent channel SNR coefficient (in the next time slot) \\
    \bottomrule
  \end{tabular}
\end{table}

The first three examples are one-step lookahead cases: one-step energy lookahead, one-step channel lookahead, and one-step joint energy-channel lookahead.
The design assumes an exact one-step lookahead, but the proposed schemes remain effective as long as the lookahead is reasonably accurate.
For simplicity, we continue to assume that both the energy arrivals and channel SNR coefficients are i.i.d.
This simplification affects only the probability weights assigned to sample paths, so it does not diminish the significance of our proposed methods in general cases.
Under the i.i.d.\ assumption, $q$ is insensitive to the one-step lookahead and hence requires no special design.

The fourth example considers a general first-order Markov model for temporal correlation in the energy arrivals.
In this case, exploiting the most recent energy arrival to accurately estimate $p$ and $q$ becomes crucial for designing an effective power control scheme.

\subsection{One-Step Energy Lookahead}\label{subsec:one_step_energy_lookahead}

Suppose that the one-step lookahead energy $\lookahead{E}_t=E_t$.
Then, the DMCR estimate $p$ used for action selection in line~\ref{line:action_selection} of Algorithm~\ref{alg:rca-rl} can be replaced by
\begin{equation}
  \lookahead{p}
  \approx \expect \left(\frac{\upperclip{E_t}{C_t}}{C_t} \middle| \lookahead{E}_t=E_t\right)
  \approx \frac{\upperclip{\lookahead{E}_t}{c}}{c}, \label{eq:estimation2_of_p}
\end{equation}
which is a conditional-expectation variant of~\eqref{eq:estimation_of_p}.
This yields the one-step energy lookahead RCA-RL scheme (RCA-RL-ELK), as shown in \suppPseudoCode.
During the first $D$ steps, $p$ is still estimated online using~\eqref{eq:estimation_of_p}, but it is used only to warm-start $q$ at $t=D$.

\subsection{One-Step Channel Lookahead}\label{subsec:one_step_channel_lookahead}

Suppose that the one-step lookahead channel SNR coefficient $\lookahead{\Gamma}_t=\Gamma_{t+1}$.
Then, the parameter $\hat{\gamma}$ can be estimated by
\begin{equation}
  \hat{\gamma}
  \approx s \lookahead{\Gamma}_t + \hat{\gamma}_0, \quad \hat{\gamma}_0,s \ge 0, \label{eq:estimation_of_gamma_hat}
\end{equation}
so we need to learn the new approximate relative-value function
\begin{equation}
  \hat{h}_{q,\hat{\gamma}_0,s}(b,\gamma)
  \geqdef \hat{h}_{q,s\gamma+\hat{\gamma}_0}(b), \label{eq:relative_value_function_approximation3}
\end{equation}
where $\hat{h}_{q,\hat{\gamma}}(b)$ is defined by~\eqref{eq:relative_value_function_approximation2}.
Using~\eqref{eq:estimation_of_gamma_hat} and~\eqref{eq:relative_value_function_approximation3}, we modify Algorithm~\ref{alg:rca-rl} to obtain the one-step channel lookahead RCA-RL scheme (RCA-RL-CLK), as shown in \suppPseudoCode.

\subsection{One-Step Joint Energy-Channel Lookahead}\label{subsec:one_step_joint_energy_channel_lookahead}

Suppose that the one-step lookahead energy and channel SNR coefficient at time $t$ are $\lookahead{E}_t=E_t$ and $\lookahead{\Gamma}_t=\Gamma_{t+1}$, respectively.
Then, using~\eqref{eq:estimation2_of_p}--\eqref{eq:relative_value_function_approximation3}, we obtain the one-step joint energy-channel lookahead RCA-RL scheme (RCA-RL-ECLK), as shown in \suppPseudoCode.

\subsection{Markov Energy Arrivals}\label{subsec:markov_energy_arrivals}

Consider a first-order time-invariant Markov process for the energy arrivals.
In this case, the available contextual information $O_t$ at time $t$ is the most recent energy arrival $E_{t-1}$, which can be used to predict the DMCR $p$ and the parameter $q$.
Then, the contextual extension of the DMCR estimate in~\eqref{eq:estimation_of_p} is given by
\begin{equation}
  p
  = \expect \left( \frac{B_{t+1}-B_t+U_t}{C_t} \middle| O_t=E_{t-1}\right), \label{eq:estimation3_of_p}
\end{equation}
and the parameter $q$ can be predicted by a parameterized function $f(E_{t-1}; \boldsymbol{\theta})$, where $\boldsymbol{\theta}$ is learned via relative-value regression of $\hat{h}_{f(E_{t-1}; \boldsymbol{\theta}),\hat{\gamma}}(b)$.
These ideas are conceptual.
For a simple and efficient design, we adopt the following approach.

First, we define the normalized context $\lookahead{O}$ as
\begin{equation}
  \lookahead{O}
  = \frac{\upperclip{O_t}{c}}{c}.\label{eq:normalized_context}
\end{equation}
There is no loss of information if $O_t \le c$.
Even when $O_t > c$, $\lookahead{O}=1$ still serves as a reasonable contextual representative for capturing the temporal correlation of energy arrivals in most scenarios.

Next, we define the binning function $\bin_K$ as
\begin{equation}
  \bin_{K}(x)
  \geqdef \min\{\floor{Kx}, K-1\}, \quad x \in [0,1],\label{eq:binning_function}
\end{equation}
respectively, where $K\ge 1$ denotes the number of bins and $\floor{x}$ denotes the largest integer less than or equal to $x$.
Accordingly, we can predict the DMCR $p$ by maintaining $K$ separate estimates $(p_k)_{k=0}^{K-1}$, where $p_k$ is updated only when $\bin_{K}(\lookahead{O})=k$.

Then, we define the predictor for $q$ as
\begin{equation}
  q
  \approx \kappa(\lookahead{O}; \boldsymbol{\theta})
  \geqdef \mlp_{[1,J,1]}(\lookahead{O}; \boldsymbol{\theta}),\label{eq:prediction_of_q}
\end{equation}
where $\mlp_{[1,J,1]}(\cdot; \boldsymbol{\theta})$ denotes a $[1,J,1]$ multilayer perceptron (MLP), with ReLU activation in the hidden layer and sigmoid activation in the output layer.

When the energy arrivals exhibit strong temporal persistence, $\lookahead{O}$ may serve as a reasonable predictor for $q$, in the spirit of the fixed fraction policy~\cite{shaviv2016universally}.
In this case, we define
\begin{equation}
  q
  \approx \bar{\kappa}(\lookahead{O}; \boldsymbol{\theta})
  \geqdef (1-\alpha)\lookahead{O} + \alpha \lookahead{q},\label{eq:prediction_of_q_with_persistence}
\end{equation}
where
\begin{equation}
  (\alpha, \lookahead{q})
  \eqdef \mlp_{[1,J,2]}(\lookahead{O}; \boldsymbol{\theta}).
\end{equation}

Finally, combining~\eqref{eq:normalized_context}--\eqref{eq:prediction_of_q_with_persistence}, we obtain two versions of the Markov-energy-arrival RCA-RL scheme, as shown in \suppPseudoCode.
The first, RCA-RL-M, uses~\eqref{eq:prediction_of_q}, while the second, RCA-RL-MP, uses~\eqref{eq:prediction_of_q_with_persistence}.

\section{Simulation Results}\label{sec:simulation_results}

In this section, we present simulation results to evaluate the performance of the proposed power control schemes.
For comparison, we also evaluate the performance of the optimal policies, which are computed via policy iteration (PI)~\cite[Sec.~8.6.1]{puterman2005markov} based on the MDP model in Sec.~\ref{sec:problem_formulation} with or without the energy/channel lookahead extensions in Sec.~\ref{sec:prediction_based_extensions}.\footnote{%
  To compute the optimal policies, we discretize the continuous state and action spaces.
  The battery level, channel SNR coefficient, and energy consumption are quantized into 250, 50, and 250 levels, respectively, for the optimal online policy without lookahead.
  For the optimal one-step energy or channel lookahead policy, battery level, channel SNR coefficient, lookahead value, and energy consumption are quantized into 150, 20, 20, and 150 levels, respectively.
  For states that do not exactly match the quantization grid, interpolation is used to determine the optimal action.
  Due to the curse of dimensionality and computational constraints, we have to exclude the optimal one-step energy-channel lookahead policy from the comparison.\label{ft:optimal}%
}
In addition, we include several online power control schemes based on Lyapunov optimization~\cite{amirnavaei2016online} and generic RL methods~\cite{mnih2015human,schulman2017proximal,fujimoto2018addressing} as baselines.
Table~\ref{tab:schemes} lists all the schemes to be compared in the simulation.

\begin{table}
  \centering
  \caption{Power Control Schemes Compared in the Simulations}\label{tab:schemes}
  \tablelinespread
  \begin{tabular}{rp{5cm}}
    \toprule
    Scheme Abbreviation &
    Description \\
    \midrule
    OPT &
    Optimal online policy (computed by PI) \\
    OCA-OL &
    Table~\ref{tab:closed_form_policies} \\
    RCA-OL &
    Table~\ref{tab:closed_form_policies} \\
    OCA-OL-SA &
    Algorithm~\ref{alg:ca-ol-a} \\
    RCA-OL-SA &
    Algorithm~\ref{alg:ca-ol-a} \\
    RCA-OLA-A &
    Algorithm~\ref{alg:rca-ola-a} \\
    RCA-RL &
    Algorithm~\ref{alg:rca-rl} \\
    Lyapunov &
    Lyapunov-optimization-based policy proposed in~\cite{amirnavaei2016online} \\
    OPT-ELK &
    Optimal one-step energy lookahead policy (computed by PI) \\
    RCA-RL-ELK &
    One-step energy lookahead RCA-RL scheme (Sec.~\ref{subsec:one_step_energy_lookahead}) \\
    OPT-CLK &
    Optimal one-step channel lookahead policy (computed by PI) \\
    RCA-RL-CLK &
    One-step channel lookahead RCA-RL scheme (Sec.~\ref{subsec:one_step_channel_lookahead}) \\
    RCA-RL-ECLK &
    One-step joint energy-channel lookahead RCA-RL scheme (Sec.~\ref{subsec:one_step_joint_energy_channel_lookahead}) \\
    RCA-RL-M &
    Markov-energy-arrival RCA-RL scheme (Sec.~\ref{subsec:markov_energy_arrivals}) \\
    RCA-RL-MP &
    RCA-RL-M variant for strong temporal persistence (Sec.~\ref{subsec:markov_energy_arrivals}) \\
    DQN &
    Deep Q-Network~\cite{mnih2015human} \\
    PPO &
    Proximal Policy Optimization~\cite{schulman2017proximal} \\
    TD3 &
    Twin Delayed Deep Deterministic Policy Gradient~\cite{fujimoto2018addressing} \\
    \bottomrule
  \end{tabular}
\end{table}

We follow the performance-evaluation framework in~\cite[Sec.~2.2.2]{yang2025power}, which is based on the following concepts:
\begin{itemize}
  \item \term{Nominal mean-to-capacity ratio} (NMCR):
  \begin{equation}
    \text{NMCR}
    \geqdef \frac{\expect E_t}{c}. \label{eq:nmcr}
  \end{equation}
  By removing the clip function, this ratio is easier to compute and use in practice than the MCR $\mcratio(P_E,c)$ (Eq.~\eqref{eq:dmcr}), where $P_E$ denotes the distribution of $E_t$.
  On the other hand, however, energy-arrival distributions with the same NMCR can have different MCRs.
  This is demonstrated by Table~\ref{tab:mcr}, which compares the MCRs of the three distribution types used in the simulation.
  For these special types, their parameters, and consequently the MCR, are uniquely determined by the NMCR.
  \item \term{Nominal signal-to-noise ratio} (NSNR) in decibels (dB):
  \begin{equation}
    \text{NSNR}
    \geqdef 10\log_{10} (\expect \Gamma_t \expect \upperclip{E_t}{c})
    = 10\log_{10} \clippedmean, \label{eq:nsnr}
  \end{equation}
  where $\clippedmean\eqdef\clippedmean(P_E,c)$ is the clipped mean defined by~\eqref{eq:clipped_mean}.
  Given the NSNR and MCR, the battery capacity $c$ can be computed by
  \begin{equation}
    c
    = \frac{\clippedmean}{\text{MCR}}
    = \frac{10^{\text{NSNR/10}}}{\text{MCR}}.
  \end{equation}
\end{itemize}

\begin{table}
  \centering
  \caption{MCRs of Bernoulli, Exponential, and Uniform Distributions \cite[Table~2.2]{yang2025power}}\label{tab:mcr}
  \tablelinespread
  \begin{tabular}{ccccc}
    \toprule
    Distribution   & MCR for NMCR $\tilde{p}$ & $\tilde{p}=0.1$ & $0.5$ & $0.9$ \\
    \midrule
    $\bernoulli_q$ & $\tilde{p}$              & $0.1$           & $0.5$ & $0.9$ \\
    $\expon_\lambda$ & $\tilde{p}(1 - \en^{-1/\tilde{p}})$
    & $0.1000$ & $0.4323$ & $0.6037$ \\[0.5ex]
    $\uniform_b$ & \(
    \begin{cases}
      \tilde{p},               & \tilde{p} \in [0,\frac{1}{2}] \\
      1-\dfrac{1}{4\tilde{p}}, & \tilde{p}>\frac{1}{2}         \\
    \end{cases}
    \) & $0.1$ & $0.5$ & $0.7222$ \\[4ex]
    \bottomrule
  \end{tabular}
\end{table}

To better visualize the simulation results, we introduce a performance metric called the \term{online multiplicative factor}.
For a power control policy $\sigma$ (not necessarily online), this metric is defined as
\begin{equation}
  \text{OMF}(\sigma)
  \geqdef \frac{\averageReward(\sigma)}{g^*}, \label{eq:online_multiplicative_factor}
\end{equation}
where $g^*$ denotes the maximum online throughput, achieved by the OPT scheme.

\subsection{Performance of Online Power Control Schemes}\label{subsec:performance_of_online_schemes}

In this subsection, we compare the performance of online power control schemes, including OCA-OL, RCA-OL, OCA-OL-SA, RCA-OL-SA, RCA-OLA-A, RCA-RL, and Lyapunov.

The basic simulation settings are summarized in Table~\ref{tab:simulation_settings}.
In particular, the throughput of a power control scheme is evaluated by the average reward over $10^3$ episodes, each consisting of $10^4$ steps.
Unless stated otherwise, these settings are used throughout this section, and any differences are noted explicitly.

\begin{table}
  \centering
  \caption{Simulation Settings}\label{tab:simulation_settings}
  \tablelinespread
  \begin{tabular}{rp{3cm}}
    \toprule
    Parameter                            & Setting                                           \\
    \midrule
    Episodes                             & $10^3$                                            \\
    Steps of each episode                & $10^4$                                            \\
    Initial battery level $B_1$          & uniform on $[0,c]$                                \\
    energy-arrival distribution          & $\bernoulli_q$, $\expon_\lambda$, or $\uniform_b$ \\
    Channel SNR coefficient distribution & $\expon_1$ (Rayleigh)                             \\
    NMCR                                 & $0.1, 0.5, 0.9$                                   \\
    NSNR                                 & $0$, $5$, $10$, \ldots, $30$ dB                   \\
    \bottomrule
  \end{tabular}
\end{table}%

The parameter settings for OCA- and RCA-based simple adaptive schemes (excluding RCA-RL) are summarized in Table~\ref{tab:oca_rca_parameters}.

\begin{table}
  \centering
  \caption{Parameter Settings for OCA- and RCA-based Simple Adaptive Schemes}\label{tab:oca_rca_parameters}
  \tablelinespread
  \begin{tabular}{rl}
    \toprule
    Parameter &
    Value \\
    \midrule
    Learning rate $\eta$ &
    $\max\{1/t,10^{-4}\}$ \\
    Update interval $T$ (for RCA-OLA-A) &
    $100$ \\
    \bottomrule
  \end{tabular}
\end{table}%

The parameter settings for RCA-RL and its extensions are summarized in Table~\ref{tab:rca_rl_parameters}.
To reliably evaluate RCA-RL and its descendant algorithms, we trained $N_{\text{run}}=10$ policies using $N_{\text{run}}$ different policy seeds and training-data seeds, and for a fixed number $S_{\text{tr}}$ of training steps.
We then froze the resulting RCA policies by fixing $q$ and $\hat{\gamma}$, and evaluated them on the same data as the other policies.
Their average performance is reported for comparison.

\begin{table}
  \centering
  \caption{Parameter Settings for RCA-RL and Its Extensions}\label{tab:rca_rl_parameters}
  \tablelinespread
  \begin{tabular}{rl}
    \toprule
    Parameter &
    Setting \\
    \midrule
    Learning rates $\eta_1$, $\eta_2$ &
    $\max\{1/t,10^{-4}\}$ \\
    Learning rate $\eta_3$ &
    $10^{-3}$ \\
    Replay buffer capacity $M$ &
    $16$ \\
    Minibatch size $N$ &
    $1$ \\
    Minibatch optimizer &
    Adam \\
    Exploration probability $\epsilon$ &
    $0$ \\
    Warm-up period $D$ &
    $100$ \\
    Target-network update interval $T_{\text{target}}$ &
    $500$ \\
    \bottomrule
  \end{tabular}
\end{table}%

The Lyapunov policy is evaluated only for the uniform energy-arrival distribution with $\text{NMCR}=0.1$, since its applicability relies on stringent constraints on the maximum charging and discharging energy per time slot.
Its parameters are set as $E_{\min}=0$, $E_{\max}=c$, $\maxDischargeEnergy=\maxChargeEnergy=0.2c$, and $\eta=0.01$, where $\maxDischargeEnergy$ and $\maxChargeEnergy$ correspond to $\Delta t P_{\max}$ and $E_{c,\max}$ in~\cite{amirnavaei2016online}, respectively.

A summary of the performance loss of OCA- and RCA-based schemes relative to the corresponding optimal online policies is provided in Table~\ref{tab:performance_loss}.
A more detailed performance comparison across different energy-arrival distributions and NMCRs, with NSNR as the swept variable, is presented in the figures in \suppFiguresA.
Further numerical results on stability, convergence, and hyperparameter sensitivity of RCA-based adaptive schemes are provided in \suppStability.
These results show that the adaptive DMCR estimate $p$ is highly stable, and that RCA-RL exhibits satisfactory stability, rapid convergence, and robust performance over a practical range of learning rates $\eta_3$.

\begin{table}
  \centering
  \caption{Percentage Performance Loss Relative to the Optimal Policies}\label{tab:performance_loss}
  \tablelinespread
  \begin{tabular}{ccc}
    \toprule
    Scheme    & Average   & Maximum   \\
    \midrule
    OCA-OL    & $1.61\%$  & $6.19\%$  \\
    RCA-OL    & $1.41\%$  & $3.65\%$  \\
    OCA-OL-SA & $2.09\%$  & $4.57\%$  \\
    RCA-OL-SA & $0.68\%$  & $1.76\%$  \\
    RCA-OLA-A & $0.53\%$  & $1.26\%$  \\
    RCA-RL ($S_{\text{tr}}=20\,000$) &
    $0.53\%$ &
    $1.46\%$ \\
    Lyapunov  & $27.32\%$ & $37.42\%$ \\
    \bottomrule
  \end{tabular}
\end{table}

Table~\ref{tab:performance_loss} shows that the proposed OCA/RCA-based schemes stay close to the optimal policies in both average and maximum loss.
Among them, RCA-OLA-A and RCA-RL perform the best overall, with an average loss of about $0.5\%$ and a maximum loss around $1\%$--$2\%$.
RCA-OL-SA also performs well, while RCA-OL, OCA-OL, and OCA-OL-SA have larger losses, especially in terms of maximum loss.
The significant improvement from RCA-OL to RCA-OL-SA indicates that the simple adaptive design based on the DMCR estimate $p$ is effective for non-Bernoulli energy-arrival distributions.
The further improvement from RCA-OL-SA to RCA-OLA-A is also notable, demonstrating the benefit of computing the maximin optimal linear-policy slope from the adaptive DMCR estimate $p$ rather than from the static MCR.

By contrast, the Lyapunov policy is much worse, with a much larger average loss and maximum loss.
For this reason, it is not plotted in the figures in \suppFiguresA\ because including it in the same plot would compress the curves of the other schemes near the top of the figure.

\subsection{Performance under Charging and Discharging Constraints}

In this subsection, we evaluate the performance of the proposed RCA-based adaptive schemes under additional constraints on the maximum charging and discharging energy per time slot.
As noted in Remark~\ref{re:modified_oca_and_rca}, the maximum-dischargeable-energy constraint can be enforced by clipping the proposed RCA-based policies.

Three schemes are evaluated: RCA-OLA-A, RCA-RL, and the Lyapunov policy, which is specifically designed to handle charging and discharging constraints.
The simulation is conducted under uniform energy arrivals with $\text{NMCR}=0.1$ and $\text{NSNR}=0$--$30\,\text{dB}$.
The per-slot maximum dischargeable energy $\maxDischargeEnergy$ is set to $0.05c$, and the per-slot maximum chargeable energy $\maxChargeEnergy$ is set to $0.04c$, $0.05c$, or $0.06c$.

Table~\ref{tab:limited_performance_loss} shows that, under uniform energy arrivals, both RCA-based schemes remain close to optimal even under the imposed charging and discharging constraints.
However, the two schemes are not equally effective: RCA-OLA-A incurs an average loss of only $0.07\%$ and a maximum loss of $0.23\%$, whereas RCA-RL incurs an average loss of $0.20\%$ and a maximum loss of $1.03\%$.
By contrast, the Lyapunov policy performs much worse under the same constraints, with an average loss exceeding $3\%$ and a maximum loss above $10\%$.
Fig.~\ref{fig:limited-uniform-mf} provides a more detailed comparison of these schemes.
A key difference between RCA-OLA-A and RCA-RL is that the former performs better at low NSNRs.

\begin{table}
  \centering
  \caption{Percentage Performance Loss Relative to the Optimal Policies Under Maximum Per-Slot Charging and Discharging Constraints for Uniform Energy Arrivals with $\text{NMCR}=0.1$ and $\text{NSNR}=0$--$30\,\text{dB}$}\label{tab:limited_performance_loss}
  \tablelinespread
  \begin{tabular}{ccc}
    \toprule
    Scheme    & Average  & Maximum   \\
    \midrule
    RCA-OLA-A & $0.07\%$ & $0.23\%$  \\
    RCA-RL ($S_{\text{tr}}=20\,000$) &
    $0.20\%$ &
    $1.04\%$ \\
    Lyapunov  & $3.34\%$ & $10.38\%$ \\
    \bottomrule
  \end{tabular}
\end{table}

\begin{figure*}
  \centering
  \includegraphics{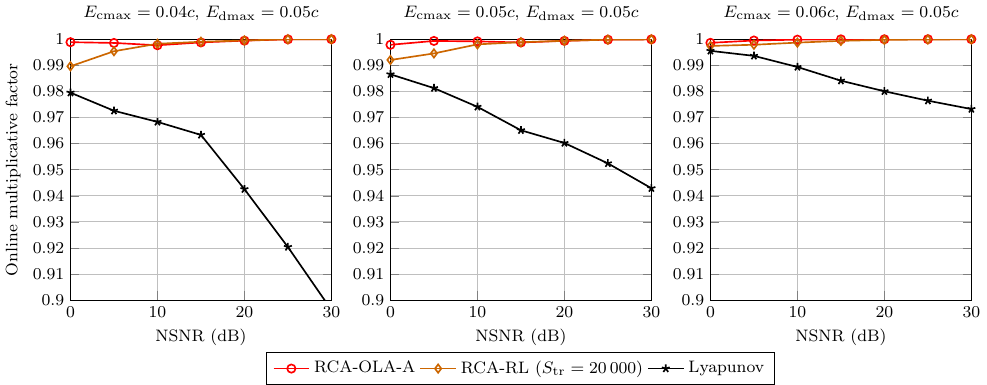}
  \caption{Online multiplicative factors of the power control schemes under uniform energy arrivals with $\text{NMCR}=0.1$ and per-slot maximum charging and discharging energy constraints.}\label{fig:limited-uniform-mf}
\end{figure*}

\subsection{Performance of RCA-RL with One-Step Lookahead}\label{subsec:performance_of_rca_rl_with_lookahead}

In this subsection, we evaluate the performance of the RCA-RL scheme with one-step lookahead.

Table~\ref{tab:lookahead_performance_loss} shows that the RCA-RL schemes with one-step energy or channel lookahead achieve average performance losses of about $0.5\%$ and maximum losses below $2\%$ relative to the corresponding optimal one-step lookahead policies.
This indicates that the proposed RCA-RL extensions can effectively exploit the additional lookahead information, thereby achieving near-optimality relative to the corresponding optimal lookahead policies.
However, RCA-RL-CLK converges much more slowly and performs poorly with only 20\,000 training steps; therefore, we set the number of training steps to 100\,000 for RCA-RL-CLK.
The same training-step setting is also used for RCA-RL-ECLK.
Since the optimal one-step joint energy-channel lookahead policies are unavailable, we evaluate the performance loss of RCA-RL-ECLK relative to the corresponding optimal one-step energy lookahead policy (OPT-ELK).
Compared with RCA-RL-ELK, RCA-RL-ECLK achieves a smaller average loss of $0.14\%$ and a smaller maximum loss of $0.95\%$, demonstrating a modest additional benefit from incorporating channel lookahead information.

\begin{table}
  \centering
  \caption{Percentage Performance Loss Relative to the Corresponding Optimal One-Step Lookahead Policies}\label{tab:lookahead_performance_loss}
  \tablelinespread
  \begin{tabular}{cccc}
    \toprule
    Scheme & Average & Maximum & Relative to \\
    \midrule
    RCA-RL-ELK ($S_{\text{tr}}=20\,000$) &
    $0.50\%$ &
    $1.77\%$ &
    OPT-ELK \\
    RCA-RL-CLK ($S_{\text{tr}}=100\,000$) &
    $0.57\%$ &
    $1.68\%$ &
    OPT-CLK \\
    RCA-RL-ECLK ($S_{\text{tr}}=100\,000$) &
    $0.14\%$ &
    $0.95\%$ &
    OPT-ELK \\
    \bottomrule
  \end{tabular}
\end{table}

As shown in the figures in \suppFiguresB, one-step energy lookahead yields a significant performance gain, whereas one-step channel lookahead provides only a small performance gain, even under the optimal policy with one-step channel lookahead.
This is reasonable because, before the next energy arrival, there is no other decision opportunity, whereas there is still one chance to schedule before the next channel SNR coefficient is realized.
In other words, the ``true'' one-step channel lookahead is the channel SNR coefficient in the current time slot, rather than in the next time slot.

Based on the above observations, we focus on the performance of RCA-RL-ELK in the rest of this subsection.
A key concern is whether RCA-RL-ELK remains effective when the energy lookahead information is imperfect, since one-step lookahead may be unreliable in practice.
To evaluate the performance of RCA-RL-ELK under imperfect lookahead information, we introduce a noisy version of the one-step energy lookahead:
\begin{equation}
  \lookahead{E}_t
  \eqdef \lowerclip{E_{t+1} + Z_t}{0}, \label{eq:noisy_energy_lookahead}
\end{equation}
where $(Z_t)_{t\ge 1}$ is an i.i.d.\ sequence of zero-mean Gaussian random variables, referred to as the \term{lookahead noise}.

\begin{figure}
  \centering
  \includegraphics{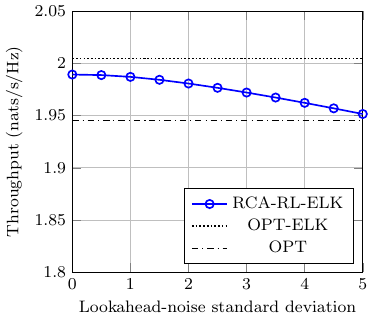}
  \caption{Throughput of RCA-RL-ELK versus lookahead-noise standard deviation in a uniform energy-arrival scenario with $\text{NMCR}=0.5$ and $\text{NSNR}=10\,\text{dB}$.}\label{fig:elk_noise}
\end{figure}

Fig.~\ref{fig:elk_noise} illustrates the robustness of RCA-RL-ELK to imperfect energy lookahead.
As the standard deviation of the lookahead noise increases, the achieved throughput decreases only mildly, from about 1.989 at zero noise to about 1.952 at the largest tested noise level.
Meanwhile, the variability across runs grows gradually, indicating that the noisy lookahead primarily introduces a small degradation rather than an abrupt performance collapse.
Even under substantial lookahead uncertainty, RCA-RL-ELK remains close to the ideal lookahead benchmark OPT-ELK and consistently outperforms the no-lookahead OPT baseline in this uniform energy-arrival scenario with $\text{NMCR}=0.5$ and $\text{NSNR}=10\,\text{dB}$ (i.e., $E_t \sim \uniform_c$ with $c=20$).

\subsection{RCA-RL versus Generic Model-Free RL}\label{subsec:rca_rl_vs_generic_rl}

In this subsection, we compare RCA-RL against several generic model-free RL methods, including DQN~\cite{mnih2015human}, Proximal Policy Optimization (PPO)~\cite{schulman2017proximal}, and Twin Delayed Deep Deterministic Policy Gradient (TD3)~\cite{fujimoto2018addressing}, an enhanced variant of DDPG.

The simulations are conducted under two settings: uniform energy arrivals with $\text{NMCR}=0.5$ and $\text{NSNR}=10\,\text{dB}$, and wind energy arrivals~\cite{ding_2021_5516539} with $\text{NMCR}=0.2$ and $\text{NSNR}=10\,\text{dB}$.
In both cases, the channel SNR coefficients are assumed to be i.i.d.\ Rayleigh.
For uniform energy arrivals, each run consists of 30\,000 steps, whereas for wind energy arrivals, only 24\,306 steps are simulated because of the limited size of the available wind-energy dataset (see Remark~\ref{re:wind_dataset}).
All policies are trained online for 20\,000 steps and then evaluated over the remaining steps until the end of each simulation run; for each setting, we perform 100 runs with different random seeds, using the same seed set for all algorithms to ensure a fair comparison.
The only exception is PPO, which stops training after 20\,480 steps because its rollout buffer size is 2048.
Therefore, for a fair comparison, all policies are evaluated after 20\,480 steps, even though the other policies stop training after 20\,000 steps.

\begin{remark}[Wind Time-Series Dataset]\label{re:wind_dataset}
  The wind time-series dataset~\cite{ding_2021_5516539} was collected from a single turbine at an inland wind farm and was first introduced in~\cite[Chap.~2]{ding2020data}.
  In our simulations, we use the 10-minute wind-power samples from this dataset.
  Although the dataset spans one year, some time instants are missing.
  We therefore treat consecutive time instants separated by gaps of at most 20 minutes (i.e., one or two samples) as belonging to the same continuous episode, and we set the battery level to zero at the beginning of each episode.
  As shown in~\cite[Tabs.~2.6--2.8]{ding2020data}, the persistence model performs best for short-term forecasting, which suggests that wind power remains relatively stable over short intervals and that ignoring missing data over such short periods is unlikely to distort the time series.
  We run each policy over all episodes of length at least 100.
  In total, these data comprise 24\,306 steps across 126 episodes, and the 20\,480th step falls in the 113th episode.
  In order to exploit the temporal correlations in the wind energy arrivals, we include the most recent energy arrival in the state information for all algorithms.
  Accordingly, in the wind-energy-arrival case, we use RCA-RL-M and RCA-RL-MP (both with the context-bin number $K=10$) instead of the original RCA-RL, since these extensions are designed to exploit the Markovian structure of the energy arrivals.
\end{remark}

The Stable-Baselines3 (SB3) implementation~\cite{raffin2021stable} (v2.7.1) is used for DQN, PPO, and TD3.
The main hyperparameters, except for the network architecture and learning rates, for these algorithms are listed in Table~\ref{tab:generic_rl_parameters}.
MLPs with ReLU activation functions in the hidden layers are used for function approximation, with the same architecture for both the policy and value networks within each algorithm, except for DQN, which has only a value network.
For each algorithm, we use the same hidden-layer architecture across the two energy-arrival settings, namely, two hidden layers with the same number of neurons in each layer.
This is a common choice for simple control problems.
The number of neurons in each layer for each algorithm was chosen through a broad parameter search to yield a reasonably good, moderate-sized network.

\begin{table}
  \centering
  \caption{Hyperparameters of DQN, PPO, and TD3, Except Network Architecture and Learning Rates}\label{tab:generic_rl_parameters}
  \tablelinespread
  \begin{tabular}{>{\centering\arraybackslash}m{2.7cm}cc>{\centering\arraybackslash}m{1.6cm}}
    \toprule
    Parameter & DQN & PPO & TD3 \\
    \midrule
    Replay or rollout buffer size &
    20\,000 &
    2048 &
    10\,000 \\
    Minibatch size &
    8 &
    64 &
    128 \\
    Warm-up steps &
    1000 &
    N/A &
    500 \\
    Target update interval &
    100 &
    N/A &
    2
    \\
    Soft update coefficient &
    0.005 &
    N/A &
    0.005 \\
    Exploration &
    SB3-default &
    SB3-default &
    normal noise ($\sigma=0.05$) \\
    Discount factor &
    0.99 &
    0.99 &
    0.99 \\
    Number of discrete actions &
    101 &
    N/A &
    N/A \\
    Optimizer &
    Adam &
    Adam &
    Adam \\
    \bottomrule
  \end{tabular}
\end{table}

The results are summarized in Tables~\ref{tab:generic_rl_comparison_under_uniform} and~\ref{tab:generic_rl_comparison_under_wind}.
For each generic RL method (DQN, PPO, and TD3), the first row reports the main configuration used for comparison, while the second row reports a larger network architecture only as a reference to show that the reported performance is not an artifact of using a small network.
To provide a comprehensive comparison, we also report the performance of the optimal policy (OPT) under uniform energy arrivals and the optimal policy under the first-order empirical distribution of wind energy arrivals, denoted as IID-OPT.

\begin{table*}
  \centering
  \caption{Performance Comparison of RCA-RL and Generic Model-Free RL Methods Under Uniform Energy Arrivals with $\text{NSNR}=10\,\text{dB}$}\label{tab:generic_rl_comparison_under_uniform}
  \tablelinespread
  \begin{tabular}{ccccc}
    \toprule
    Algorithm & Network Architecture & Learning Rate & \multicolumn{2}{c}{Evaluation-Period Throughput (nats/s/Hz)} \\
    \cmidrule(lr){4-5}
    & (tuned for $\text{NMCR}=0.5$) & & $\text{NMCR}=0.5$ & $\text{NMCR}=0.1$ \\
    \midrule
    DQN &
    $[2,6,6,101]$ value network &
    0.003 &
    1.887 &
    1.890 \\
    &
    $[2,24,24,101]$ value network &
    0.001 &
    1.890 &
    1.883 \\
    PPO &
    $[2,6,6,1]$ policy and value networks &
    0.01 &
    1.915 &
    1.957 \\
    &
    $[2,24,24,1]$ policy and value networks &
    0.003 &
    1.917 &
    1.953 \\
    TD3 &
    $[2,18,18,1]$ policy and value networks &
    0.003 &
    1.901 &
    1.969 \\
    &
    $[2,72,72,1]$ policy and value networks &
    0.003 &
    1.902 &
    1.962 \\
    \textbf{RCA-RL} &
    $[1,1]$ value network (Eq.~\eqref{eq:relative_value_function_approximation2}) &
    0.001 &
    \textbf{1.934} &
    \textbf{2.050} \\
    OPT &
    N/A &
    N/A &
    1.945 (asymptotic) &
    2.059 (asymptotic) \\
    \bottomrule
  \end{tabular}
\end{table*}

\begin{table*}
  \centering
  \caption{Performance Comparison of RCA-RL and Generic Model-Free RL Methods Under Wind Energy Arrivals with $\text{NMCR}=0.2$ and $\text{NSNR}=10\,\text{dB}$}\label{tab:generic_rl_comparison_under_wind}
  \tablelinespread
  \begin{tabular}{>{\centering\arraybackslash}m{2cm}>{\centering\arraybackslash}m{7cm}cc}
    \toprule
    Algorithm & Network Architecture & Learning Rate & Evaluation-Period Throughput (nats/s/Hz) \\
    \midrule
    DQN &
    $[3,6,6,101]$ value network &
    0.001 &
    1.817 \\
    &
    $[3,24,24,101]$ value network &
    0.0003 &
    1.816 \\
    PPO &
    $[3,6,6,1]$ policy and value networks &
    0.003 &
    1.832 \\
    &
    $[3,24,24,1]$ policy and value networks &
    0.003 &
    1.832 \\
    TD3 &
    $[3,18,18,1]$ policy and value networks &
    0.001 &
    1.811 \\
    &
    $[3,72,72,1]$ policy and value networks &
    0.0003 &
    1.816 \\
    \textbf{RCA-RL-M} ($K=10$) &
    $[1,1]$ value network (Eq.~\eqref{eq:relative_value_function_approximation2}) with a $[1,1,1]$ network for predicting $q$ (Eq.~\eqref{eq:prediction_of_q} with $J=1$) &
    0.001 &
    \textbf{1.865} \\
    \textbf{RCA-RL-MP} ($K=10$) &
    $[1,1]$ value network (Eq.~\eqref{eq:relative_value_function_approximation2}) with a roughly $[1,1,2,1]$ network for predicting $q$ (Eq.~\eqref{eq:prediction_of_q_with_persistence} with $J=1$) &
    0.001 &
    \textbf{1.872} \\
    IID-OPT &
    N/A &
    N/A &
    1.824 \\
    \bottomrule
  \end{tabular}
\end{table*}

Figs.~\ref{fig:generic_rl_comparison_under_uniform} and~\ref{fig:generic_rl_comparison_under_wind} illustrate the corresponding throughput learning curves.
For uniform energy arrivals, each run is partitioned into 100-step windows, whereas for wind energy arrivals, episodes serve as windows of variable duration.
In both scenarios, we compute the window-averaged reward for each run and report the mean across 100 independent runs as the throughput.
The shaded regions indicate the interquartile range across runs.

\begin{figure*}
  \centering
  \includegraphics{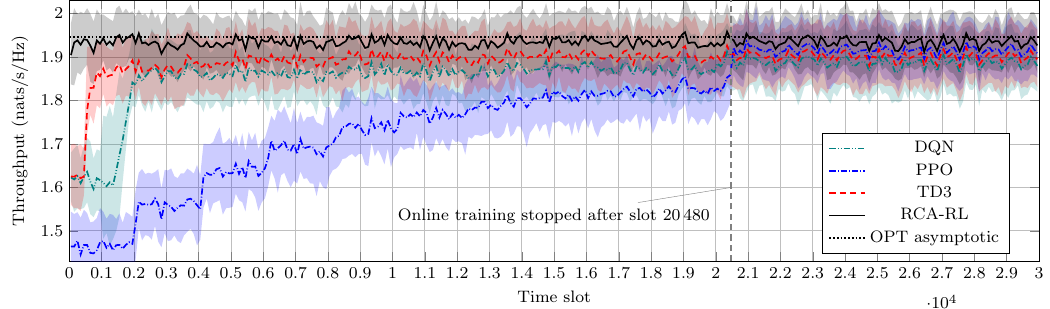}
  \caption{Throughput of RCA-RL and generic model-free RL methods under uniform energy arrivals with $\text{NMCR}=0.5$ and $\text{NSNR}=10\,\text{dB}$.}\label{fig:generic_rl_comparison_under_uniform}
\end{figure*}

\begin{figure*}
  \centering
  \includegraphics{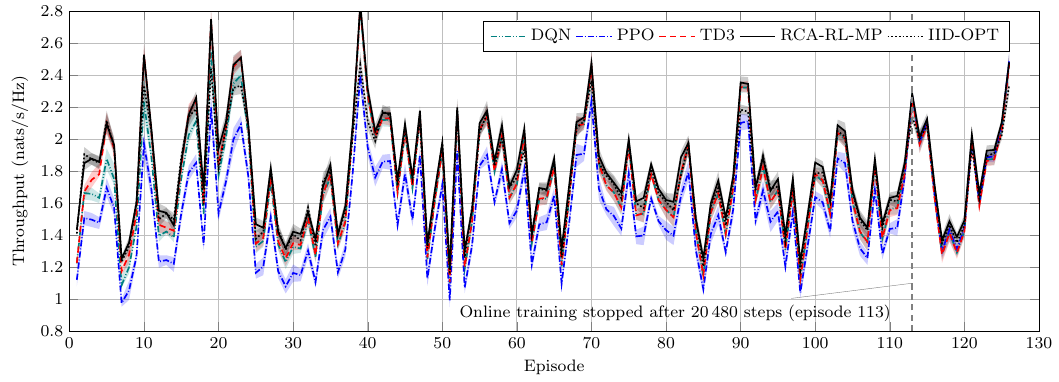}
  \caption{Throughput of RCA-RL-MP and generic model-free RL methods under wind energy arrivals with $\text{NMCR}=0.2$ and $\text{NSNR}=10\,\text{dB}$.}\label{fig:generic_rl_comparison_under_wind}
\end{figure*}

Overall, Tables~\ref{tab:generic_rl_comparison_under_uniform} and~\ref{tab:generic_rl_comparison_under_wind}, together with Figs.~\ref{fig:generic_rl_comparison_under_uniform} and~\ref{fig:generic_rl_comparison_under_wind}, show that the proposed RCA-RL and its extensions for Markov energy arrivals consistently outperform the generic model-free RL baselines in both the uniform- and wind-energy settings.
In the uniform case ($\text{NMCR}=0.5$), RCA-RL achieves a throughput of 1.934, which is close to the optimal value of 1.945.
In the wind-energy case, RCA-RL-M and RCA-RL-MP achieve throughputs of 1.865 and 1.872, respectively, substantially exceeding those of the generic RL baselines as well as the IID-OPT benchmark of 1.824.
Note that, among the three RL methods, only PPO achieves a throughput of 1.832, slightly exceeding that of IID-OPT.
This indicates that RCA-RL-M and RCA-RL-MP are more effective than generic RL methods in exploiting temporal correlations.
Furthermore, as shown in Figs.~\ref{fig:generic_rl_comparison_under_uniform} and~\ref{fig:generic_rl_comparison_under_wind}, RCA-RL and its extensions converge much faster than the three RL methods, which is consistent with their low-dimensional value function approximation.

Note that Table~\ref{tab:generic_rl_comparison_under_uniform} also reports the performance of all schemes under $\text{NMCR}=0.1$.
These results further highlight an important distinction between RCA-RL and generic RL methods.
The latter are tuned under $\text{NMCR}=0.5$, and although they perform reasonably well in that setting, their performance deteriorates noticeably when the operating condition changes to $\mathrm{NMCR}=0.1$.
This suggests that generic RL methods may be highly sensitive to the underlying system parameters, and that even extensive hyperparameter tuning may, in some cases, fail to achieve satisfactory performance across different i.i.d.\ scenarios.
In contrast, as demonstrated in Sec.~\ref{subsec:performance_of_online_schemes}, RCA-RL with a fixed parameter setting (Table~\ref{tab:rca_rl_parameters}) performs consistently well across a range of i.i.d.\ scenarios and in the wind-energy-arrival scenario.
Such robustness is difficult to achieve with generic RL methods and may even be impractical without substantial retuning.

\section{Conclusion}\label{sec:discussion_and_conclusion}

We propose a linear-policy-based approximation to the relative-value function, two parameterized clipped affine policies, and their corresponding closed-form policies and adaptive schemes for power control in point-to-point EH wireless communication systems.
The low complexity and high performance of these algorithms are demonstrated through comparative analysis in Tables~\ref{tab:comparison} and~\ref{tab:rl_overhead_comparison} and extensive simulation results in Sec.~\ref{sec:simulation_results}.
These results suggest that the proposed power control schemes, along with their underling design approach, form a competitive and promising building block for practical EH wireless communication systems.

In particular, we establish a general design approach for energy management in EH wireless communication systems, which consists of the following steps:
\begin{enumerate}
  \item \textbf{Tractable future-value approximation}: Identify an analytically tractable approximation of the relative-value function under a simplified model setting.
  This approximation should capture the essential structure of the optimal policy while remaining simple enough to enable analytical policy improvement, or at least efficient numerical optimization or a closed-form approximation.
  \item \textbf{Robust current-action optimization}: Apply worst-case analysis to determine the best action for the current time slot under the full model, using the identified approximation of the relative-value function.
  This yields a policy that is robust to model uncertainties under certain assumed statistics.
  \item \textbf{Adaptive parameter learning}: Develop an adaptive scheme to learn the parameters of the relative-value function approximation and the policy from online data, ensuring stable and sample-efficient learning.
\end{enumerate}

The main challenge in the area of EH communication energy management lies in the interplay of energy-storage dynamics under EH settings and various application-specific factors (constraints or utilities).
Incorporating all such factors simultaneously into the MDP model typically leads to an analytically intractable problem.
The success of this work shows that, as long as the battery-limited EH constraint is the dominant performance bottleneck, it is sufficient to predict future values using a simplified model that captures only the basic energy-storage dynamics under EH settings, while other factors can be effectively handled through worst-case analysis in the current-action optimization step.
In the problem addressed in this work, these other factors include varying channel quality and the per-slot maximum-dischargeable-energy constraint, both of which are ignored in the prediction of future values.

Besides, this work also highlights the importance of incorporating domain knowledge into the design of RL algorithms for EH communication energy management.
As shown in Sec.~\ref{subsec:rca_rl_vs_generic_rl}, the proposed RCA-RL scheme and its extensions for Markov energy arrivals, designed by exploiting the approximate structure of the optimal policy and the relative-value function, significantly outperform generic model-free RL methods that do not leverage such domain knowledge, in both performance and convergence speed.
This suggests that, for complex control problems in EH communication systems, it is often more effective to design RL algorithms that are informed by the underlying problem structure rather than relying solely on generic RL methods.
Nevertheless, due to issues related to stability, convergence, and hyperparameter sensitivity, RL-based approaches are still not a fully satisfactory solution for EH communication energy management; developing non-RL approaches will be the focus of our future work.


\appendices

\section{Proofs of Theorems~\ref{th:onepoint_bellman2_solution}--\ref{th:onepoint_bellman2_solution2}}\label{app:bellman2_solution}

\begin{proofof}{Theorem~\ref{th:onepoint_bellman2_solution}}
  It suffices to verify that $(r(e), h_1)$ satisfies~\eqref{eq:bellman2} for $E \sim \onepoint{e}$.
  We have
  \begin{align*}
    f(u)
    &\eqdef r(u) + \expect_{E \sim \onepoint{e}} h_1(\upperclip{b-u+E}{c}) \\
    &= r(u) + h_1(\upperclip{b-u+e}{c}) \\
    &= r(u) + r(e) + r'(e)(\upperclip{b-u+e}{c} - e) \\
    &=
    \begin{cases}
      r(u) + r(e) + r'(e)(c-e), &\text{$u \in [0,b+e-c)$}, \\
      r(u) + r(e) + r'(e)(b-u), &\text{$u \in [b+e-c,b]$},
    \end{cases}
  \end{align*}
  which implies that $f(u)$ is strictly increasing on $[0,\upperclip{b}{e})$ and is strictly decreasing on $(\upperclip{b}{e}, b]$.
  Therefore, the action $u=\upperclip{b}{e}=\clippedGreedyPolicy[e](b)$ is optimal, and
  \begin{align*}
    \sup_{u \in [0,b]} f(u)
    &= f(\upperclip{b}{e}) \\
    &\eqvar{(a)} r(\upperclip{b}{e}) + r(e) + r'(e)(b - \upperclip{b}{e}) \\
    &= r(e) + h_1(b),
  \end{align*}
  where (a) follows from $\upperclip{b}{e} \ge b+e-c$.
\end{proofof}

\begin{proofof}{Theorem~\ref{th:bernoulli_bellman2_solution}}
  It suffices to verify that $(ph_2(c), h_2)$ satisfies~\eqref{eq:bellman2} for $E \sim \bernoulli_p$.
  We have
  \begin{align}
    &\sup_{u\in [0,b]} (r(u) + \expect_{E \sim \bernoulli_p} h_2(\upperclip{b-u+E}{c})) \notag \\
    &= \sup_{u\in [0,b]} [r(u) + ph_2(c) + (1-p)h_2(b-u)] \notag \\
    &= ph_2(c) + \sup_{u\in [0,b]} \left[r(u) + \sup_{\substack{(u_i)_{i=1}^\infty: u_i\ge 0,\\ \sum_{i=1}^\infty u_i \le b-u}} \sum_{i=1}^\infty (1-p)^i r(u_i)\right] \label{eq:bernoulli_bellman2_solution.optimal_action} \\
    &= ph_2(c) + \sup_{\substack{(u_i)_{i=0}^\infty: u_i\ge 0,\\ \sum_{i=0}^\infty u_i \le b}} \sum_{i=0}^\infty (1-p)^i r(u_i) \notag \\
    &= ph_2(c) + h_2(b). \notag
  \end{align}
  By~\cite[Thm.~3 and its proof as well as Thm.~6]{yang2020maximin}, the optimal $u_0$ in~\eqref{eq:bernoulli_relative_value_function}, or equivalently, the optimal action $u$ in~\eqref{eq:bernoulli_bellman2_solution.optimal_action}, is given by $u_0 = \maximinOptimalPolicy[p](x)$.
  Then, we have
  \begin{align*}
    h_2(x)
    &= r(\maximinOptimalPolicy[p](x)) + (1-p)h_2(x-\maximinOptimalPolicy[p](x)) \\
    &= r(\maximinOptimalPolicy[p](x)) + (1-p)h_2(\reserve{\maximinOptimalPolicy[p]}(x)) \\
    &= \sum_{i=0}^{\tilde{M}(x) - 1} (1-p)^i r(\maximinOptimalPolicy[p](\reserve{\maximinOptimalPolicy[p]}^{(i)}(x))),
  \end{align*}
  where the last equality follows from
  \begin{align*}
    &\reserve{\maximinOptimalPolicy[p]}^{(\tilde{M}(x))}(x) \\
    &= \reserve{\maximinOptimalPolicy[p]}^{(\tilde{M}(x)-1)}(x) - \maximinOptimalPolicy[p](\reserve{\maximinOptimalPolicy[p]}^{(\tilde{M}(x)-1)}(x)) \\
    &= \reserve{\maximinOptimalPolicy[p]}^{(\tilde{M}(x)-1)}(x) - \reserve{\maximinOptimalPolicy[p]}^{(\tilde{M}(x)-1)}(x)
    = 0,
  \end{align*}
  because
  \begin{align*}
    1
    &\le \tilde{M}(\reserve{\maximinOptimalPolicy[p]}^{(\tilde{M}(x)-1)}(x)) \\
    &\le \tilde{M}(\reserve{\maximinOptimalPolicy[p]}^{(\tilde{M}(x)-2)}(x)) - 1 \\
    &\cdots \\
    &\le \tilde{M}(x) - (\tilde{M}(x) - 1)
    = 1,
  \end{align*}
  which is also true for the degenerate case $\tilde{M}(x)=1$.
  Note that $\tilde{M}(\reserve{\maximinOptimalPolicy[p]}(x)) \le \tilde{M}(x) - 1$ for $\tilde{M}(x) \ge 2$, because
  \(
  [1+p(\reserve{\maximinOptimalPolicy[p]}(x)+\tilde{M}(x)-1)](1-p)^{\tilde{M}(x)-1}
  < 1.
  \)
\end{proofof}

\begin{proofof}{Theorem~\ref{th:onepoint_bellman2_solution2}}
  For $b=b_0=e/q$, we have
  \begin{align*}
    f(u)
    &\eqdef r(u) + \expect\hat{h}_q(\upperclip{b-u+E}{c}) \\
    &= r(u) + \frac{1}{q} r(q\upperclip{b_0-u+e}{c}) \\
    &=
    \begin{dcases}
      r(u) + \frac{1}{q} r(qc), &\text{$u \in [0,b_0+e-c)$}, \\
      r(u) + \frac{1}{q} r(q(b_0-u+e)), &\text{$u \in [b_0+e-c,b]$}.
    \end{dcases}
  \end{align*}
  Then,
  \[
    f'(u)
    =
    \begin{dcases}
      r'(u), &\text{$u \in [0,b_0+e-c)$}, \\
      r'(u) - r'(e-q(u-e)), &\text{$u \in [b_0+e-c,b]$},
    \end{dcases}
  \]
  which implies that $f(u)$ is strictly increasing on $[0,e)$ and is strictly decreasing on $(e, b_0]$.
  Therefore, the action $u = e = \sigma(b_0)$ is optimal, and
  \[
    \sup_{u \in [0,b_0]} f(u)
    = f(e)
    = r(e) + \hat{h}_q(b_0).
  \]
  It is also clear that
  \[
    \phi^{(n)}(x)
    = (1-q)^n(x-b_0) + b_0
    \quad \text{for all $n \ge 0$},
  \]
  which implies that $\lim_{n \to \infty} \phi^{(n)}(x)
  = b_0$.
\end{proofof}

\section{Proofs of Theorems~\ref{th:optimistic_clipped_affine_policy} and~\ref{th:robust_clipped_affine_policy}}\label{app:clipped_affine_policy}

\begin{proofof}{Theorem~\ref{th:optimistic_clipped_affine_policy}}
  Let
  \[
    f(u)
    \eqdef r(\gamma u) + \hat{h}_{q,\hat{\gamma}}(b-u+\upperclip{e}{c-b+u}).
  \]
  For $u \in [0, b_0(e))$,
  \[
    f(u)
    = r(\gamma u) + \hat{h}_{q,\hat{\gamma}}(c)
  \]
  is increasing in $u$, so the maximum of $f(u)$ can always be attained at some point in $[b_0(e),b]$.

  For $u \in [b_0(e),b]$,
  \[
    f(u)
    = r(\gamma u) + \hat{h}_{q,\hat{\gamma}}(b-u+e).
  \]
  If $\gamma = 0$, then $f(u)$ is decreasing on $[b_0(e),b]$ and thus attains its maximum at $u = b_0(e) = \optimisticClippedAffinePolicy[e,q,\hat{\gamma}](b,0)$.
  If $\gamma > 0$, then
  \begin{equation}
    f'(u)
    = g(u)
    \eqdef \dfrac{\gamma}{1+\gamma u} - \frac{\hat{\gamma}}{1+\hat{\gamma}q(b-u+e)} \label{eq:optimistic_waterfilling}
  \end{equation}
  is strictly decreasing in $u$.
  Therefore, $f(u)$ is strictly concave on $[b_0(e),b]$.
  It is easy to see that
  \[
    b_1
    \eqdef \frac{q(b+e)-1/\gamma+1/\hat{\gamma}}{1+q}
  \]
  is the unique solution to $g(u) = 0$ over $\real$.

  If $b_1 \in [b_0(e),b]$, then $f(u)$ attains its maximum at $u = b_1$;
  otherwise, we have $b_1 > b$ or $b_1 < b_0(e)$.
  If $b_1 > b$, then $f'(u)$ is positive on $[b_0(e),b]$, hence $f(u)$ is strictly increasing on $[b_0(e),b]$, and therefore $f(u)$ attains its maximum at $u = b = \upperclip{b_1}{b}$.
  If $b_1 < b_0(e)$, then $f'(u)$ is negative on $[b_0(e),b]$, hence $f(u)$ is strictly decreasing on $[b_0(e),b]$, and therefore $f(u)$ attains its maximum at $u = b_0(e) = \lowerclip{b_1}{b_0(e)}$.
  In summary, the optimal action is given by $\clip{b_1}{b_0(e)}{b}$.
\end{proofof}

\begin{proofof}{Theorem~\ref{th:robust_clipped_affine_policy}}
  Let
  \[
    f(u)
    \eqdef r(\gamma u) + (1-p) \hat{h}_{q,\hat{\gamma}}(b-u) + p \hat{h}_{q,\hat{\gamma}}(c).
  \]
  If $\gamma = 0$, then $f(u)$ is strictly decreasing on $[0,b]$, and hence the optimal action is $u = 0 = \robustClippedAffinePolicy[p,q,\hat{\gamma}](b,0)$.

  Next, we suppose that $\gamma > 0$.
  Then,
  \begin{equation}
    f'(u)
    = g(u)
    \eqdef \dfrac{\gamma}{1+\gamma u} - \frac{(1-p)\hat{\gamma}}{1+\hat{\gamma}q(b-u)} \label{eq:robust_waterfilling}
  \end{equation}
  is strictly decreasing in $u$.
  Therefore, $f(u)$ is strictly concave on $[0,b]$.
  It is easy to see that
  \[
    b_0
    \eqdef \frac{qb-(1-p)/\gamma+1/\hat{\gamma}}{1-p+q}
  \]
  is the unique solution to $g(u) = 0$ over $\real$.

  If $b_0 \in [0,b]$, then $f(u)$ attains its maximum at $u = b_0$;
  otherwise, we have $b_0 > b$ or $b_0 < 0$.
  If $b_0 > b$, then $f'(u)$ is positive on $[0,b]$, hence $f(u)$ is strictly increasing on $[0,b]$, and therefore $f(u)$ attains its maximum at $u = b = \upperclip{b_0}{b}$.
  If $b_0 < 0$, then $f'(u)$ is negative on $[0,b]$, hence $f(u)$ is strictly decreasing on $[0,b]$, and therefore $f(u)$ attains its maximum at $u = 0 = \lowerclip{b_0}{0}$.
  In summary, the optimal action is given by $\clip{b_0}{0}{b}$.
\end{proofof}

\section*{Conflict of Interest}

The authors S.~Yang and H.~Wu are inventors of Chinese Patent ZL202510318380.1 (CN119854923B), which relates to a power control method described in this paper and is assigned to Zhejiang Gongshang University.
All other authors declare no conflict of interest.

\bibliographystyle{IEEEtran}
\bibliography{IEEEabrv,eh6-1}

\begin{thebibliography}{10}
\providecommand{\url}[1]{#1}
\csname url@samestyle\endcsname
\providecommand{\newblock}{\relax}
\providecommand{\bibinfo}[2]{#2}
\providecommand{\BIBentrySTDinterwordspacing}{\spaceskip=0pt\relax}
\providecommand{\BIBentryALTinterwordstretchfactor}{4}
\providecommand{\BIBentryALTinterwordspacing}{\spaceskip=\fontdimen2\font plus
\BIBentryALTinterwordstretchfactor\fontdimen3\font minus
  \fontdimen4\font\relax}
\providecommand{\BIBforeignlanguage}[2]{{%
\expandafter\ifx\csname l@#1\endcsname\relax
\typeout{** WARNING: IEEEtran.bst: No hyphenation pattern has been}%
\typeout{** loaded for the language `#1'. Using the pattern for}%
\typeout{** the default language instead.}%
\else
\language=\csname l@#1\endcsname
\fi
#2}}
\providecommand{\BIBdecl}{\relax}
\BIBdecl

\bibitem{wu2026robust}
H.~Wu, S.~Yang, H.~Gao, D.~Wang, J.~Chen, and G.~Yu, ``Robust clipped affine
  policy for online power control in energy harvesting communications,'' in
  \emph{2026 {{IEEE International Symposium}} on {{Information Theory}}
  ({{ISIT}})}.\hskip 1em plus 0.5em minus 0.4em\relax Guangzhou, China: IEEE,
  Jun. 2026, pp. 1--6.

\bibitem{ulukus2015energy}
S.~Ulukus, A.~Yener, E.~Erkip, O.~Simeone, M.~Zorzi, P.~Grover, and K.~Huang,
  ``Energy harvesting wireless communications: A review of recent advances,''
  \emph{{IEEE} J. Sel. Areas Commun.}, vol.~33, no.~3, pp. 360--381, Mar. 2015.

\bibitem{ku2016advances}
M.-L. Ku, W.~Li, Y.~Chen, and K.~J. Ray~Liu, ``Advances in energy harvesting
  communications: Past, present, and future challenges,'' \emph{{IEEE} Commun.
  Surveys Tuts.}, vol.~18, no.~2, pp. 1384--1412, 2016.

\bibitem{ma2020sensing}
D.~Ma, G.~Lan, M.~Hassan, W.~Hu, and S.~K. Das, ``Sensing, computing, and
  communications for energy harvesting {IoTs}: A survey,'' \emph{{IEEE} Commun.
  Surveys Tuts.}, vol.~22, no.~2, pp. 1222--1250, 2020.

\bibitem{hu2020modeling}
S.~Hu, X.~Chen, W.~Ni, X.~Wang, and E.~Hossain, ``Modeling and analysis of
  energy harvesting and smart grid-powered wireless communication networks: A
  contemporary survey,'' \emph{{IEEE} Trans. Green Commun. Netw.}, vol.~4,
  no.~2, pp. 461--496, Jun. 2020.

\bibitem{alamu2025machine}
O.~Alamu, T.~O. Olwal, and E.~M. Migabo, ``Machine learning applications in
  energy harvesting internet of things networks: A review,'' \emph{{IEEE}
  Access}, vol.~13, pp. 4235--4266, 2025.

\bibitem{yang2025power}
S.~Yang and J.~Chen, ``Power control for battery-limited energy harvesting
  communications,'' \emph{Foundations and Trends® in Communications and
  Information Theory}, vol.~22, no. 2-3, pp. 185--393, 2025.

\bibitem{kazerouni2015optimal}
A.~Kazerouni and A.~\"Ozg\"ur, ``Optimal online strategies for an energy
  harvesting system with {Bernoulli} energy recharges,'' in \emph{Proc. 2015
  13th International Symposium on Modeling and Optimization in Mobile, Ad Hoc,
  and Wireless Networks ({WiOpt})}.\hskip 1em plus 0.5em minus 0.4em\relax
  Mumbai, India: IEEE, May 2015, pp. 235--242.

\bibitem{shaviv2015capacity}
D.~Shaviv and A.~\"Ozg\"ur, ``Capacity of the {AWGN} channel with random
  battery recharges,'' in \emph{Proc. 2015 {IEEE} International Symposium on
  Information Theory ({ISIT})}.\hskip 1em plus 0.5em minus 0.4em\relax Hong
  Kong, Hong Kong: IEEE, Jun. 2015, pp. 136--140.

\bibitem{shaviv2016universally}
------, ``Universally near optimal online power control for energy harvesting
  nodes,'' \emph{{IEEE} J. Sel. Areas Commun.}, vol.~34, no.~12, pp.
  3620--3631, Dec. 2016.

\bibitem{yang2020maximin}
S.~Yang and J.~Chen, ``A maximin optimal online power control policy for energy
  harvesting communications,'' \emph{{IEEE} Trans. Wireless Commun.}, vol.~19,
  no.~10, pp. 6708--6720, Oct. 2020.

\bibitem{zibaeenejad2022optimal}
A.~Zibaeenejad, S.~Yang, and J.~Chen, ``On optimal power control for energy
  harvesting communications with lookahead,'' \emph{{IEEE} Trans. Wireless
  Commun.}, vol.~21, no.~6, pp. 4054--4067, Jun. 2022.

\bibitem{wang2021optimality}
Y.~Wang, A.~Zibaeenejad, Y.~Jing, and J.~Chen, ``On the optimality of the
  greedy policy for battery limited energy harvesting communications,''
  \emph{{IEEE} Trans. Inf. Theory}, vol.~67, no.~10, pp. 6548--6563, Oct. 2021.

\bibitem{yang2020.6maximin}
S.~Yang and J.~Chen, ``A maximin optimal online power control policy for energy
  harvesting communications,'' in \emph{Proc. {ICC} 2020 - 2020 {IEEE}
  International Conference on Communications ({ICC})}.\hskip 1em plus 0.5em
  minus 0.4em\relax Dublin, Ireland: IEEE, Jun. 2020, pp. 1--6.

\bibitem{garmaroudi2022linear}
\BIBentryALTinterwordspacing
H.~M. Garmaroudi, Z.~Dou, S.~Yang, and J.~Chen, ``On linear power control
  policies for energy harvesting communications,'' 2022. [Online]. Available:
  \url{https://arxiv.org/abs/2207.10230}
\BIBentrySTDinterwordspacing

\bibitem{arafa2018online}
A.~Arafa, A.~Baknina, and S.~Ulukus, ``Online fixed fraction policies in energy
  harvesting communication systems,'' \emph{{IEEE} Trans. Wireless Commun.},
  vol.~17, no.~5, pp. 2975--2986, May 2018.

\bibitem{khajepour2021optimal}
A.~Khajepour and A.~Zibaeenejad, ``Optimal power control policy for fading
  channels with {Bernoulli} harvested energy,'' in \emph{2021 {{Iran Workshop}}
  on {{Communication}} and {{Information Theory}} ({{IWCIT}})}.\hskip 1em plus
  0.5em minus 0.4em\relax Tehran, Iran, Islamic Republic of: IEEE, May 2021,
  pp. 1--6.

\bibitem{ortiz2016reinforcement}
A.~Ortiz, H.~{Al-Shatri}, X.~Li, T.~Weber, and A.~Klein, ``Reinforcement
  learning for energy harvesting point-to-point communications,'' in \emph{2016
  {{IEEE International Conference}} on {{Communications}} ({{ICC}})}.\hskip 1em
  plus 0.5em minus 0.4em\relax Kuala Lumpur, Malaysia: IEEE, May 2016, pp.
  1--6.

\bibitem{masadeh2018reinforcement}
A.~Masadeh, Z.~Wang, and A.~E. Kamal, ``Reinforcement learning exploration
  algorithms for energy harvesting communications systems,'' in \emph{2018
  {{IEEE International Conference}} on {{Communications}} ({{ICC}})}.\hskip 1em
  plus 0.5em minus 0.4em\relax Kansas City, MO: IEEE, May 2018, pp. 1--6.

\bibitem{kim2018actionbounding}
H.~Kim, H.~Yang, Y.~Kim, and J.~Lee, ``Action-bounding for reinforcement
  learning in energy harvesting communication systems,'' in \emph{2018 {{IEEE
  Global Communications Conference}} ({{GLOBECOM}})}.\hskip 1em plus 0.5em
  minus 0.4em\relax Abu Dhabi, United Arab Emirates: IEEE, Dec. 2018, pp. 1--7.

\bibitem{li2019deep}
M.~Li, X.~Zhao, H.~Liang, and F.~Hu, ``Deep reinforcement learning optimal
  transmission policy for communication systems with energy harvesting and
  adaptive mqam,'' \emph{{IEEE} Trans. Veh. Technol.}, vol.~68, no.~6, pp.
  5782--5793, Jun. 2019.

\bibitem{masadeh2019actorcritic}
A.~Masadeh, Z.~Wang, and A.~E. Kamal, ``An actor-critic reinforcement learning
  approach for energy harvesting communications systems,'' in \emph{2019 28th
  {{International Conference}} on {{Computer Communication}} and {{Networks}}
  ({{ICCCN}})}.\hskip 1em plus 0.5em minus 0.4em\relax Valencia, Spain: IEEE,
  Jul. 2019, pp. 1--6.

\bibitem{qiu2019deep}
C.~Qiu, Y.~Hu, Y.~Chen, and B.~Zeng, ``Deep deterministic policy gradient
  ({DDPG})-based energy harvesting wireless communications,'' \emph{{IEEE}
  Internet Things J.}, vol.~6, no.~5, pp. 8577--8588, Oct. 2019.

\bibitem{kim2021shallow}
H.~Kim, J.~Lee, W.~Shin, and H.~V. Poor, ``Shallow reinforcement learning for
  energy harvesting communications with imperfect channel knowledge,''
  \emph{{IEEE} J. Sel. Topics Signal Process.}, vol.~15, no.~5, pp. 1258--1271,
  Aug. 2021.

\bibitem{ozel2011transmission}
O.~Ozel, K.~Tutuncuoglu, J.~Yang, S.~Ulukus, and A.~Yener, ``Transmission with
  energy harvesting nodes in fading wireless channels: Optimal policies,''
  \emph{{IEEE} J. Sel. Areas Commun.}, vol.~29, no.~8, pp. 1732--1743, Sep.
  2011.

\bibitem{baknina2016optimal}
A.~Baknina and S.~Ulukus, ``Optimal and near-optimal online strategies for
  energy harvesting broadcast channels,'' \emph{{IEEE} J. Sel. Areas Commun.},
  vol.~34, no.~12, pp. 3696--3708, Dec. 2016.

\bibitem{baknina2018energy}
------, ``Energy harvesting multiple access channels: Optimal and near-optimal
  online policies,'' vol.~66, no.~7, pp. 2904--2917, Jul. 2018.

\bibitem{yao2023multiple}
Y.~Yao, Y.~Chen, H.~Yao, Z.~Ni, and M.~Motani, ``Multiple task resource
  allocation considering {QoS} in energy harvesting systems,'' \emph{{IEEE}
  Internet Things J.}, vol.~10, no.~9, pp. 7893--7908, May 2023.

\bibitem{zhang2026federated}
K.~Zhang, X.~Cao, and K.~B. Letaief, ``Federated learning with energy
  harvesting devices: An {MDP} framework,'' \emph{{IEEE} Trans. Wireless
  Commun.}, vol.~25, pp. 10\,886--10\,903, 2026.

\bibitem{cho2025multiagent}
H.~Cho, H.~Kim, J.~Na, S.-C. Lim, and H.~Lee, ``Multiagent distributed {DQN}
  and transfer learning for energy-efficient power management in solar
  energy-harvested small-cell networks,'' \emph{{IEEE} Internet Things J.},
  vol.~12, no.~12, pp. 19\,590--19\,604, Jun. 2025.

\bibitem{amirnavaei2016online}
F.~Amirnavaei and M.~Dong, ``Online power control optimization for wireless
  transmission with energy harvesting and storage,'' \emph{{IEEE} Trans.
  Wireless Commun.}, vol.~15, no.~7, pp. 4888--4901, Jul. 2016.

\bibitem{ku2021neuralnetworkbased}
M.-L. Ku and T.-J. Lin, ``Neural-network-based power control prediction for
  solar-powered energy harvesting communications,'' \emph{{IEEE} Internet
  Things J.}, vol.~8, no.~16, pp. 12\,983--12\,998, Aug. 2021.

\bibitem{wu2026clipped}
\BIBentryALTinterwordspacing
H.~Wu, S.~Yang, H.~Gao, D.~Wang, J.~Chen, and G.~Yu, ``Clipped affine policy:
  Low-complexity near-optimal online power control for energy harvesting
  communications over fading channels,'' 2026. [Online]. Available:
  \url{https://arxiv.org/abs/2601.07622}
\BIBentrySTDinterwordspacing

\bibitem{tse2005fundamentals}
D.~Tse and P.~Viswanath, \emph{Fundamentals of Wireless Communication}.\hskip
  1em plus 0.5em minus 0.4em\relax Cambridge, UK: Cambridge University Press,
  2005.

\bibitem{arapostathis1993discretetime}
A.~Arapostathis, V.~S. Borkar, E.~{Fern{\'a}ndez-Gaucherand}, M.~K. Ghosh, and
  S.~I. Marcus, ``Discrete-time controlled {Markov} processes with average cost
  criterion: A survey,'' \emph{SIAM Journal on Control and Optimization},
  vol.~31, no.~2, pp. 282--344, Mar. 1993.

\bibitem{hernandez-lerma2003markov}
O.~{Hern{\'a}ndez-Lerma} and J.-B. Lasserre, \emph{Markov Chains and Invariant
  Probabilities}, ser. Progress in Mathematics.\hskip 1em plus 0.5em minus
  0.4em\relax Basel ; Boston: Birkh{\"a}user, 2003, no. v. 211.

\bibitem{tutuncuoglu2012optimum}
K.~Tutuncuoglu and A.~Yener, ``Optimum transmission policies for battery
  limited energy harvesting nodes,'' \emph{{IEEE} Trans. Wireless Commun.},
  vol.~11, no.~3, pp. 1180--1189, Mar. 2012.

\bibitem{sutton2018reinforcement}
R.~S. Sutton and A.~G. Barto, \emph{Reinforcement Learning: An Introduction},
  2nd~ed., ser. Adaptive Computation and Machine Learning Series.\hskip 1em
  plus 0.5em minus 0.4em\relax Cambridge, MA: The MIT Press, 2018.

\bibitem{yang2025power.patent}
S.~Yang and H.~Wu, ``A power control method for energy harvesting wireless
  communication systems,'' Chinese Patent CN119\,854\,923B, Jun. 6, 2025.

\bibitem{kingma2017adam}
\BIBentryALTinterwordspacing
D.~P. Kingma and J.~Ba, ``Adam: A method for stochastic optimization,'' 2017.
  [Online]. Available: \url{https://arxiv.org/abs/1412.6980}
\BIBentrySTDinterwordspacing

\bibitem{puterman2005markov}
M.~L. Puterman, \emph{Markov Decision Processes: Discrete Stochastic Dynamic
  Programming}, ser. Wiley Series in Probability and Statistics.\hskip 1em plus
  0.5em minus 0.4em\relax Hoboken, NJ: Wiley-Interscience, 2005.

\bibitem{mnih2015human}
V.~Mnih, K.~Kavukcuoglu, D.~Silver, A.~A. Rusu, J.~Veness, M.~G. Bellemare,
  A.~Graves, M.~Riedmiller, A.~K. Fidjeland, G.~Ostrovski, S.~Petersen,
  C.~Beattie, A.~Sadik, I.~Antonoglou, H.~King, D.~Kumaran, D.~Wierstra,
  S.~Legg, and D.~Hassabis, ``Human-level control through deep reinforcement
  learning,'' \emph{Nature}, vol. 518, no. 7540, pp. 529--533, Feb. 2015.

\bibitem{schulman2017proximal}
\BIBentryALTinterwordspacing
J.~Schulman, F.~Wolski, P.~Dhariwal, A.~Radford, and O.~Klimov, ``Proximal
  policy optimization algorithms,'' 2017. [Online]. Available:
  \url{https://arxiv.org/abs/1707.06347}
\BIBentrySTDinterwordspacing

\bibitem{fujimoto2018addressing}
\BIBentryALTinterwordspacing
S.~Fujimoto, H.~van Hoof, and D.~Meger, ``Addressing function approximation
  error in actor-critic methods,'' in \emph{Proceedings of the 35th
  International Conference on Machine Learning}, ser. Proceedings of Machine
  Learning Research, J.~Dy and A.~Krause, Eds., vol.~80.\hskip 1em plus 0.5em
  minus 0.4em\relax PMLR, 10--15 Jul 2018, pp. 1587--1596. [Online]. Available:
  \url{https://proceedings.mlr.press/v80/fujimoto18a.html}
\BIBentrySTDinterwordspacing

\bibitem{ding_2021_5516539}
\BIBentryALTinterwordspacing
Y.~Ding, ``Wind time series dataset,'' Sep. 2021. [Online]. Available:
  \url{https://doi.org/10.5281/zenodo.5516539}
\BIBentrySTDinterwordspacing

\bibitem{ding2020data}
------, \emph{Data Science for Wind Energy}.\hskip 1em plus 0.5em minus
  0.4em\relax Erscheinungsort nicht ermittelbar: Taylor \& Francis, 2020.

\bibitem{raffin2021stable}
\BIBentryALTinterwordspacing
A.~Raffin, A.~Hill, A.~Gleave, A.~Kanervisto, M.~Ernestus, and N.~Dormann,
  ``Stable-baselines3: Reliable reinforcement learning implementations,''
  \emph{Journal of Machine Learning Research}, vol.~22, no. 268, pp. 1--8,
  2021. [Online]. Available: \url{http://jmlr.org/papers/v22/20-1364.html}
\BIBentrySTDinterwordspacing

\end{thebibliography}

\section{Algorithm Pseudocode for the RCA-RL Extensions in Section~\ref{sec:prediction_based_extensions}}
\raggedbottom

\begin{algorithm}[H]
  \caption{RCA-RL-ELK: One-Step Energy Lookahead RCA-RL Scheme}\label{alg:rca-rl-elk}
  \begin{algorithmic}[1]
    \State Hyperparameters: learning rates $\eta_1, \eta_2, \eta_3 > 0$, replay buffer capacity $M \ge 1$, minibatch size $N \ge 1$, exploration probability $\epsilon \in [0,1)$, warm-up period $D \ge 1$, and target-network update interval $T_{\text{target}} > 0$
    \Statex \Comment{Omitted; same as lines~2--6 of Algorithm~\ref{alg:rca-rl}.}
    \State Observe the initial state $(B,\Gamma, \lookahead{E})$, where $\lookahead{E}$ denotes the one-step energy lookahead
    \For{$t=1,2,3,\ldots$}
      \State $\lookahead{p} \gets \upperclip{\lookahead{E}}{c}/c$
      \State Take action $\tilde{U} \sim (1-\epsilon)
      \onepoint{U} + \epsilon \uniform_B$, where
      \(
      U \eqdef \robustClippedAffinePolicy[\lookahead{p},q,\hat{\gamma}](B,\Gamma)
      \)
      \State \algIndentParagraph{Observe the reward $R$ and the next state $(B',\Gamma', \lookahead{E}')$}
      \State \algIndentParagraph{Push $(B,R,B')$ into $\mathcal{M}$ and pop its oldest entry if the buffer size exceeds $M$}
      \State $\Delta \gets R - \hat{g} + \hat{h}_{\tilde{q},\tilde{\gamma}}(B') - \hat{h}_{\tilde{q},\tilde{\gamma}}(B)$
      \State $\hat{g} \gets \hat{g} + \eta_2 \Delta$
      \If{$t \le D$}
        \State $E \gets B' - B + \tilde{U}$, $C \gets c - B + \tilde{U}$
        \State $p \gets p + \eta_1 (E/C - p)$
        \If{$t = D$}
          \State $q \gets p$
        \EndIf
      \Else
        \Statex \Comment{Omitted; same as lines~19--24 of Algorithm~\ref{alg:rca-rl}.}
      \EndIf
      \State $(B,\Gamma,\lookahead{E}) \gets (B',\Gamma',\lookahead{E}')$
    \EndFor
  \end{algorithmic}
\end{algorithm}

\begin{algorithm}[H]
  \caption{RCA-RL-CLK: One-step Channel Lookahead RCA-RL Scheme}\label{alg:rca-rl-clk}
  \begin{algorithmic}[1]
    \State Hyperparameters: learning rates $\eta_1, \eta_2, \eta_3 > 0$, replay buffer capacity $M \ge 1$, minibatch size $N \ge 1$, exploration probability $\epsilon \in [0,1)$, warm-up period $D \ge 1$, and target-network update interval $T_{\text{target}} > 0$
    \Statex \Comment{Omitted; same as lines~2--4 of Algorithm~\ref{alg:rca-rl}.}
    \State Initialize online relative-value function $\hat{h}_{q,\hat{\gamma}_0, s}(b,\gamma)$ with parameters, e.g., $(q,\hat{\gamma}_0,s) \gets (0.5,\alpha\expect \Gamma,1-\alpha)$ with $\alpha\in (0,1)$
    \State Initialize target relative-value function $\hat{h}_{\tilde{q},\tilde{\gamma}_0,\tilde{s}}(b,\gamma)$ with parameters, e.g., $(\tilde{q},\tilde{\gamma}_0,\tilde{s}) \gets (1,\expect \Gamma,0)$
    \State Observe the initial state $(B,\Gamma,\lookahead{\Gamma})$, where $\lookahead{\Gamma}$ denotes the one-step channel lookahead
    \For{$t=1,2,3,\ldots$}
      \State $\hat{\gamma} \gets s\lookahead{\Gamma}+\hat{\gamma}_0$
      \State Take action $\tilde{U} \sim (1-\epsilon)
      \onepoint{U} + \epsilon \uniform_B$, where
      \(
      U \eqdef \robustClippedAffinePolicy[p,q,\hat{\gamma}](B,\Gamma)
      \)
      \State \algIndentParagraph{Observe the reward $R$ and the next state $(B',\Gamma',\lookahead{\Gamma}')$}
      \State \algIndentParagraph{Push $(B,\Gamma,R,B',\Gamma')$ into $\mathcal{M}$ and pop its oldest entry if the buffer size exceeds $M$}
      \Statex \Comment{Omitted; same as lines~12--13 of Algorithm~\ref{alg:rca-rl}.}
      \State $\Delta \gets R - \hat{g} + \hat{h}_{\tilde{q},\tilde{\gamma}_0,\tilde{s}}(B',\Gamma') - \hat{h}_{\tilde{q},\tilde{\gamma}_0,\tilde{s}}(B,\Gamma)$
      \State $\hat{g} \gets \hat{g} + \eta_2 \Delta$
      \If{$t = D$}
        \State $q \gets p$
      \ElsIf
          {$t > D$}
        \State \algIndentParagraph[3em]{Sample random minibatch of $N$ entries $(B_i, \Gamma_i, R_i, B_i', \Gamma_i')$ from $\mathcal{M}$}
        \State $H_i \gets R_i - \hat{g} + \hat{h}_{\tilde{q},\tilde{\gamma}_0,\tilde{s}}(B_i',\Gamma_i'), \quad i=1,\ldots,N$
        \State Perform a gradient descent step on
        \[
          L(q,\hat{\gamma}_0,s)
          \eqdef \frac{1}{N\hat{g}^2} \sum_{i=1}^N \left(H_i-\hat{h}_{q,\hat{\gamma}_0,s}(B_i,\Gamma_i)\right)^2
        \]
        \algIndent\algIndent%
        with the learning rate $\eta_3$
        \If{$(t - D) \bmod T_{\text{target}} = 0$}
          \State $(\tilde{q}, \tilde{\gamma}_0, \tilde{s}) \gets (q,\hat{\gamma}_0,s)$
        \EndIf
      \EndIf
      \State $(B,\Gamma,\lookahead{\Gamma}) \gets (B',\Gamma',\lookahead{\Gamma}')$
    \EndFor
  \end{algorithmic}
\end{algorithm}

\begin{algorithm}[H]
  \caption{RCA-RL-ECLK: One-step Joint Energy-Channel Lookahead RCA-RL Scheme}\label{alg:rca-rl-eclk}
  \begin{algorithmic}[1]
    \State Hyperparameters: learning rates $\eta_1, \eta_2, \eta_3 > 0$, replay buffer capacity $M \ge 1$, minibatch size $N \ge 1$, exploration probability $\epsilon \in [0,1)$, warm-up period $D \ge 1$, and target-network update interval $T_{\text{target}} > 0$
    \Statex \Comment{Omitted; same as lines~2--4 of Algorithm~\ref{alg:rca-rl}.}
    \Statex \Comment{Omitted; same as lines~2--3 of Algorithm~\ref{alg:rca-rl-clk}.}
    \State Observe the initial state $(B,\Gamma,\lookahead{E},\lookahead{\Gamma})$, where the pair $(\lookahead{E},\lookahead{\Gamma})$ denotes the one-step energy-channel lookahead
    \For{$t=1,2,3,\ldots$}
      \State $\lookahead{p} \gets \upperclip{\lookahead{E}}{c}/c$
      \State $\hat{\gamma} \gets s\lookahead{\Gamma}+\hat{\gamma}_0$
      \State Take action $\tilde{U} \sim (1-\epsilon)
      \onepoint{U} + \epsilon \uniform_B$, where
      \(
      U \eqdef \robustClippedAffinePolicy[\lookahead{p},q,\hat{\gamma}](B,\Gamma)
      \)
      \State \algIndentParagraph{Observe the reward $R$ and the next state $(B',\Gamma',\lookahead{E}',\lookahead{\Gamma}')$}
      \State \algIndentParagraph{Push $(B,\Gamma,R,B',\Gamma')$ into $\mathcal{M}$ and pop its oldest entry if the buffer size exceeds $M$}
      \State $\Delta \gets R - \hat{g} + \hat{h}_{\tilde{q},\tilde{\gamma}_0,\tilde{s}}(B',\Gamma') - \hat{h}_{\tilde{q},\tilde{\gamma}_0,\tilde{s}}(B,\Gamma)$
      \State $\hat{g} \gets \hat{g} + \eta_2 \Delta$
      \If{$t \le D$}
        \Statex \Comment{Omitted; same as lines~11--15 of Algorithm~\ref{alg:rca-rl-elk}.}
      \Else
        \Statex \Comment{Omitted; same as lines~15--20 of Algorithm~\ref{alg:rca-rl-clk}.}
      \EndIf
      \State $(B,\Gamma,\lookahead{E},\lookahead{\Gamma}) \gets (B',\Gamma',\lookahead{E}',\lookahead{\Gamma}')$
    \EndFor
  \end{algorithmic}
\end{algorithm}

\begin{algorithm}[H]
  \caption{RCA-RL-M and RCA-RL-MP: Markov-energy-arrival RCA-RL Schemes}\label{alg:rca-rl-markov}
  \begin{algorithmic}[1]
    \State Hyperparameters: learning rates $\eta_1, \eta_2, \eta_3 > 0$, replay buffer capacity $M \ge 1$, minibatch size $N \ge 1$, exploration probability $\epsilon \in [0,1)$, warm-up period $D \ge 1$, target-network update interval $T_{\text{target}} > 0$, and context-bin number $K \ge 1$
    \State Initialize DMCR estimates $p_0, p_1, \ldots, p_{K-1} \in [0,1]$
    \Statex \Comment{Omitted; same as lines~3--4 of Algorithm~\ref{alg:rca-rl}.}
    \State Initialize the parameters $(\boldsymbol{\theta},\hat{\gamma})$ of the online relative-value function $\hat{h}_{f(o;\boldsymbol{\theta}),\hat{\gamma}}(b)$, where
    \[
      f(o;\boldsymbol{\theta})
      \eqdef
      \begin{cases}
        \kappa(o;\boldsymbol{\theta}) &\text{for RCA-RL-M}\\
        \bar{\kappa}(o;\boldsymbol{\theta}) &\text{for RCA-RL-MP}
      \end{cases}
    \]
    \State Initialize the parameters $(\tilde{\boldsymbol{\theta}},\tilde{\gamma})$ of the target relative-value function $\hat{h}_{f(o;\tilde{\boldsymbol{\theta}}),\tilde{\gamma}}(b)$
    \State Observe the initial state $(B,\Gamma, O)$, where $O$ denotes the most recent energy arrival
    \State $\lookahead{O} \gets \upperclip{O}{c}/c$
    \For{$t=1,2,3,\ldots$}
      \State $k \gets \bin_K(\lookahead{O})$
      \State $q \gets f(\lookahead{O};\boldsymbol{\theta})$
      \State Take action $\tilde{U} \sim (1-\epsilon)
      \onepoint{U} + \epsilon \uniform_B$, where
      \(
      U \eqdef \robustClippedAffinePolicy[p_k,q,\hat{\gamma}](B,\Gamma)
      \)
      \State \algIndentParagraph{Observe the reward $R$ and the next state $(B',\Gamma', O')$}
      \State $\lookahead{O}' \gets \upperclip{O'}{c}/c$
      \State \algIndentParagraph{Push $(B,\lookahead{O},R,B',\lookahead{O}')$ into $\mathcal{M}$ and pop its oldest entry if the buffer size exceeds $M$}
      \State $E \gets B' - B + \tilde{U}$, $C \gets c - B + \tilde{U}$
      \State $p_k \gets p_k + \eta_1 (E/C - p_k)$
      \State $\Delta \gets R - \hat{g} + \hat{h}_{f(\lookahead{O}';\tilde{\boldsymbol{\theta}}),\tilde{\gamma}}(B') - \hat{h}_{f(\lookahead{O};\tilde{\boldsymbol{\theta}}),\tilde{\gamma}}(B)$
      \State $\hat{g} \gets \hat{g} + \eta_2 \Delta$
      \If{$t > D$}
        \State \algIndentParagraph[3em]{Sample random minibatch of $N$ entries $(B_i, \lookahead{O}_i, R_i, B'_i, \lookahead{O}'_i)$ from $\mathcal{M}$}
        \State $H_i \gets R_i - \hat{g} + \hat{h}_{f(\lookahead{O}'_i;\tilde{\boldsymbol{\theta}}),\tilde{\gamma}}(B_i'), \quad i=1,\ldots,N$
        \State Perform a gradient descent step on
        \[
          L(\boldsymbol{\theta},\hat{\gamma})
          \eqdef \frac{1}{N\hat{g}^2} \sum_{i=1}^N \left(H_i-\hat{h}_{f(\lookahead{O}_i;\boldsymbol{\theta}),\hat{\gamma}}(B_i)\right)^2
        \]
        \algIndent\algIndent%
        with the learning rate $\eta_3$
        \If{$(t - D) \bmod T_{\text{target}} = 0$}
          \State $(\tilde{\boldsymbol{\theta}}, \tilde{\gamma}) \gets (\boldsymbol{\theta},\hat{\gamma})$
        \EndIf
      \EndIf
      \State $(B,\Gamma,\lookahead{O}) \gets (B',\Gamma',\lookahead{O}')$
    \EndFor
  \end{algorithmic}
\end{algorithm}

\clearpage

\section{Supplementary Figures for Section~\ref{subsec:performance_of_online_schemes}}

\begin{figure}[H]
  \centering
  \includegraphics{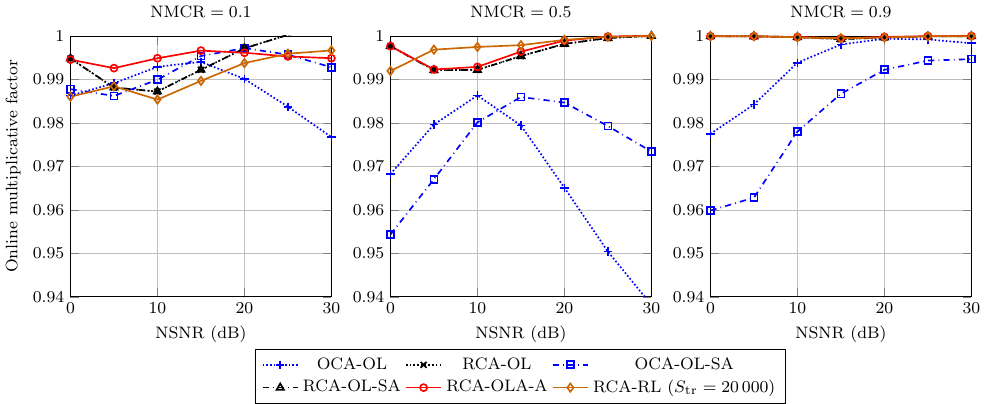}
  \caption{Online multiplicative factors of the power control schemes under Bernoulli energy arrivals.}\label{fig:bern-mf}
\end{figure}

\begin{figure}[H]
  \centering
  \includegraphics{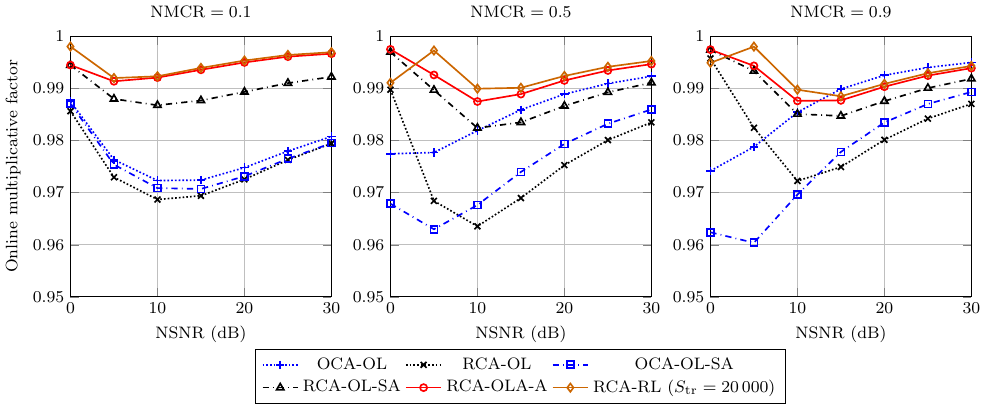}
  \caption{Online multiplicative factors of the power control schemes under exponential energy arrivals.}\label{fig:expon-mf}
\end{figure}

\begin{figure}[H]
  \centering
  \includegraphics{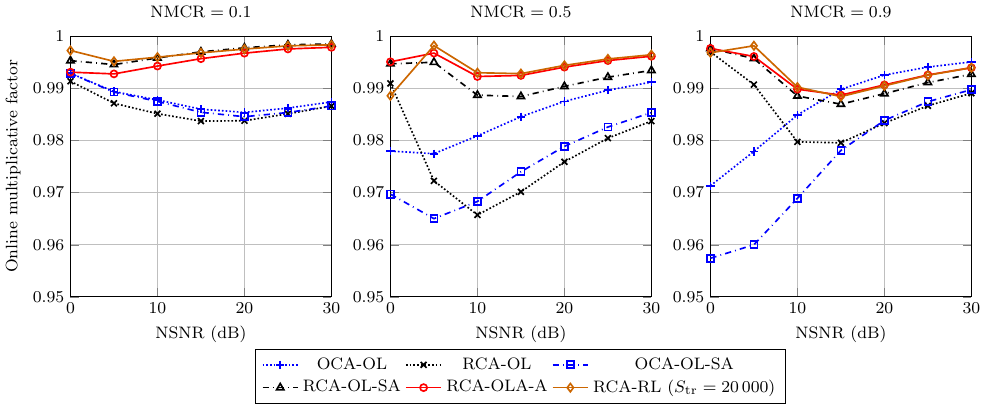}
  \caption{Online multiplicative factors of the power control schemes under uniform energy arrivals.}\label{fig:uniform-mf}
\end{figure}

\clearpage

\section{Algorithm Stability, Convergence, and Hyperparameter Sensitivity}

In this subsection, we examine the stability, convergence, and hyperparameter sensitivity of the adaptive RCA schemes.
For each plot in this subsection, we divide each run into non-overlapping windows of 100 slots and compute the window average for each tracked quantity within each run.
We then aggregate the resulting window averages across 100 independent runs with different random seeds, using the mean for the reward and the median for the other quantities.
The shaded regions indicate the interquartile range across runs for each corresponding quantity.

We first examine the evolution of the DMCR estimate $p$ in the adaptive RCA methods, namely RCA-OL-SA, RCA-OLA-A, and RCA-RL.
Fig.~\ref{fig:rcad-dynamics} shows that, for RCA with fixed $(q,\hat{\gamma})$, the estimated DMCR $p$ quickly converges to a stable value of about $0.64$ under different fixed $(q,\hat{\gamma})$ settings.
This simulation is conducted under a uniform energy-arrival distribution with $\text{NMCR}=0.5$ and $\text{NSNR}=5\,\text{dB}$.

\begin{figure}[H]
  \centering
  \includegraphics{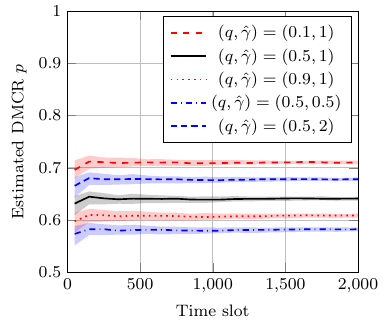}
  \caption{Evolution of the estimated DMCR of the RCA policy under different fixed $(q,\hat{\gamma})$ settings in a uniform energy-arrival scenario with $\text{NMCR}=0.5$ and $\text{NSNR}=5\,\text{dB}$.}\label{fig:rcad-dynamics}
\end{figure}

Next, we examine the stability and convergence of the RCA-RL scheme.
Fig.~\ref{fig:rcarl-dynamics} shows that, across all simulated scenarios, the RCA-RL throughput quickly converges (within $10^3$ time slots) to a stable value close to the maximum throughput achieved by the optimal policy.
The parameters $q$ and $\hat{\gamma}$ also converge to stable values.
The shaded bands indicate the variability of the estimates over time; $q$ remains relatively stable, whereas $\hat{\gamma}$ fluctuates more, but still within an acceptable range.

\begin{figure}[H]
  \centering
  \includegraphics{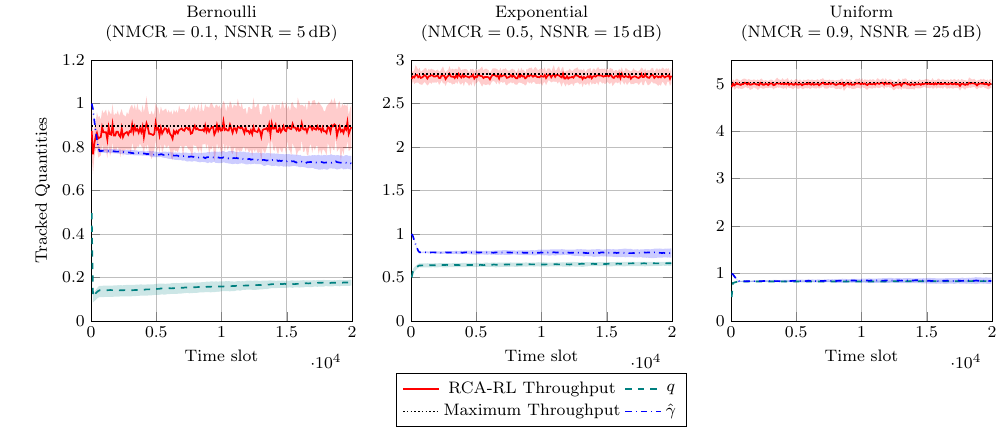}
  \caption{Evolution of the RCA-RL throughput and the parameters $q$ and $\hat{\gamma}$ under Bernoulli, exponential, and uniform energy-arrival scenarios.}\label{fig:rcarl-dynamics}
\end{figure}

Finally, we examine the sensitivity of the RCA-RL scheme to its main hyperparameter, $\eta_3$, which controls the learning rate of the relative-value function approximation.
Fig.~\ref{fig:rcarl-lr} shows that the RCA-RL scheme is relatively robust to the choice of $\eta_3$ over a reasonable range.
In terms of throughput, all three learning rates, $\eta_3 \in \set{10^{-4}, 10^{-3}, 10^{-2}}$, achieve reasonable performance, with $\eta_3 = 10^{-3}$ performing best among the tested values.
When $\eta_3$ is too small (e.g., $10^{-4}$), learning is slow, and the throughput takes longer to converge to its optimal value.
When $\eta_3$ is too large (e.g., $10^{-2}$), learning becomes unstable; although this is not immediately obvious from the throughput curve, it is reflected in the parameter estimates, especially $\hat{\gamma}$, which exhibits larger fluctuations and fails to converge to a stable value.

\begin{figure}[H]
  \centering
  \includegraphics{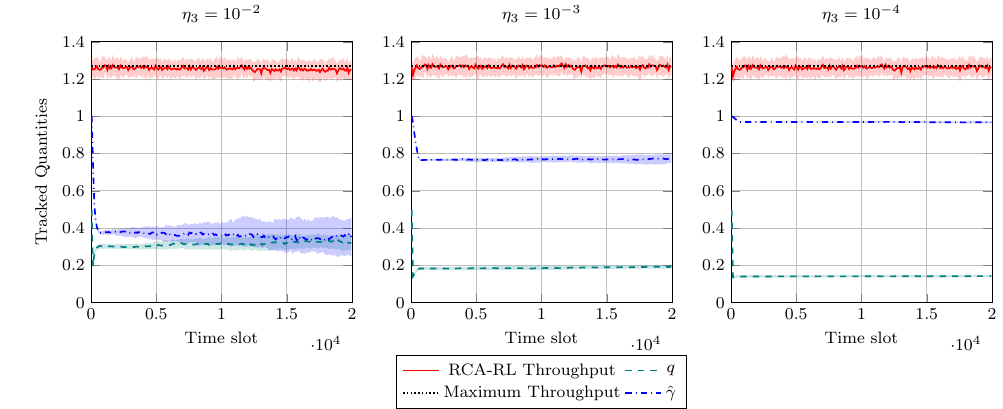}
  \caption{Sensitivity of the RCA-RL scheme to the learning rate $\eta_3$ in a uniform energy-arrival scenario with $\text{NMCR}=0.1$ and $\text{NSNR}=5\,\text{dB}$.}\label{fig:rcarl-lr}
\end{figure}

In summary, the results support the following conclusions:
\begin{itemize}
  \item The adaptive estimate of the DMCR, $p$, remains highly stable across all experiments, indicating that the proposed adaptive RCA schemes can use it as a reliable and consistent proxy for the DMCR.
  \item The RCA-RL scheme is likewise stable and converges rapidly: its throughput reaches a near-optimal plateau within about $10^3$ slots, while the learned parameters $q$ and $\hat{\gamma}$ settle to steady values, with only modest residual variation in $\hat{\gamma}$.
  \item The hyperparameter study shows that RCA-RL is reasonably robust to the choice of learning rate over a practical range, although an excessively large $\eta_3$ can destabilize the parameter estimates.
\end{itemize}
The observed stability, fast convergence, and robustness of RCA-RL are expected, since it performs low-dimensional value evaluation by learning only the two scalar parameters $(q,\hat{\gamma})$, while policy improvement is handled analytically through the RCA structure rather than through high-dimensional policy search.
Consequently, it is generally more sample-efficient and stable than generic model-free RL methods, while still exhibiting the usual learning-rate trade-off between speed and stability.

\clearpage

\section{Supplementary Figures for Section~\ref{subsec:performance_of_rca_rl_with_lookahead}}

\begin{figure}[H]
  \centering
  \includegraphics{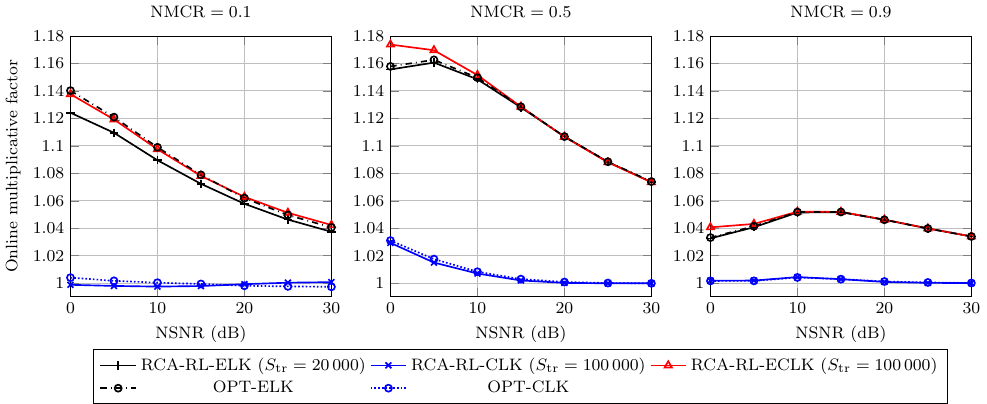}
  \caption{Online multiplicative factors of the RCA-RL scheme with one-step lookahead under Bernoulli energy arrivals.}\label{fig:lk-bern-mf}
\end{figure}

\begin{figure}[H]
  \centering
  \includegraphics{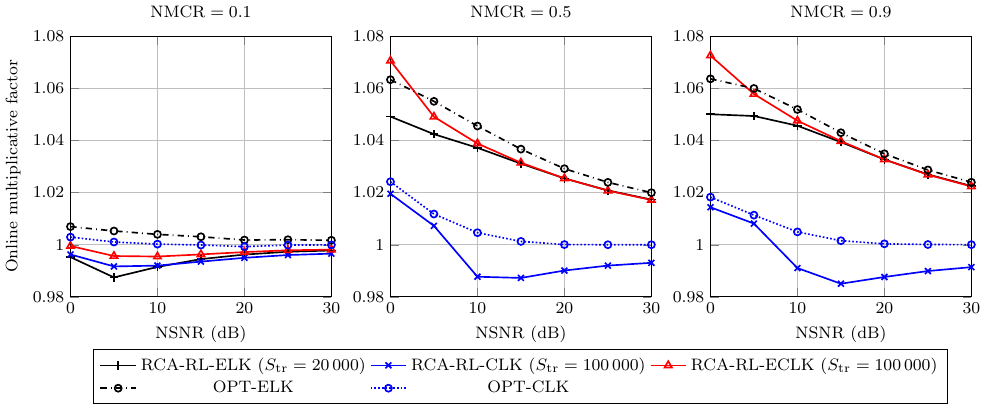}
  \caption{Online multiplicative factors of the RCA-RL scheme with one-step lookahead under exponential energy arrivals.}\label{fig:lk-expon-mf}
\end{figure}

\begin{figure}[H]
  \centering
  \includegraphics{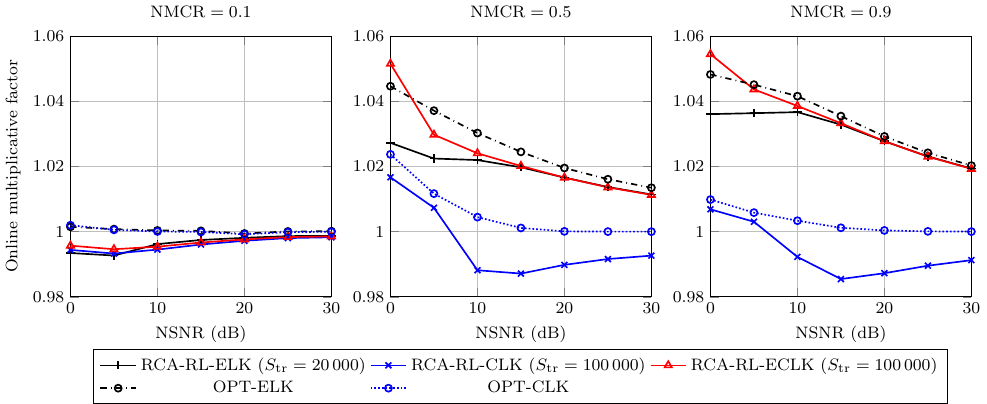}
  \caption{Online multiplicative factors of the RCA-RL scheme with one-step lookahead under uniform energy arrivals.}\label{fig:lk-uniform-mf}
\end{figure}
  \renewcommand{\thepage}{S-\arabic{page}}%
  \markboth{Supplementary Materials for Clipped Affine Policy: Omitted Appendices}{}%
}{}

\end{document}